%% file: FINAL.tex
\documentclass[lettersize,journal]{IEEEtran}
\IEEEoverridecommandlockouts
\usepackage{cite}
\usepackage[cmex10]{amsmath}
\usepackage{amsthm}
\usepackage{amssymb}
\usepackage{balance}
\usepackage[final]{graphicx}
\usepackage{color}
\usepackage{epsfig}
\usepackage{epstopdf}
\usepackage{tikz}
\usepackage{comment}
\usepackage{subfiles}
\usetikzlibrary{positioning}
\usepackage[ruled,vlined,linesnumbered,noend]{algorithm2e}

\newtheorem{lemma}{Lemma}

\newtheorem{remark}{Remark}

\newtheorem{corollary}{Corollary}

\newcommand{\bs}[1]{\boldsymbol{#1}}
\newcommand{\mc}[1]{\mathcal{#1}}

\newcommand{\mb}[1]{\mathbf{#1}}
\newcommand{\mr}[1]{\mathrm{#1}}

\newcommand{\tr}{\mathrm{Tr}}

\newcommand{\lr}[1]{\langle #1 \rangle}
\newcommand{\blr}[1]{\big\langle #1 \big\rangle}
\newcommand{\bblr}[1]{\bigg\langle #1 \bigg\rangle}
\newcommand{\Blr}[1]{\Big\langle #1 \Big\rangle}

\DeclareMathOperator*{\argmax}{arg\;max}

\definecolor{LayerColor}{RGB}{230, 230, 230}
\definecolor{InOutColor}{RGB}{240, 243, 255}
\definecolor{cellColor}{RGB}{230, 230, 230}

\begin{document}
	
	\title{Variational Bayesian Inference for Time-Varying Massive MIMO Channels: Estimation and Detection
	
	\author{Sajjad Nassirpour, \emph{Member, IEEE}, Toan-Van Nguyen, \emph{Member, IEEE}, and Duy H. N. Nguyen, \emph{Senior Member, IEEE}


\thanks{S. Nassirpour, T. V. Nguyen, and D. H. N. Nguyen are with the Department of Electrical and Computer
Engineering, San Diego State University, San Diego, CA 92182, USA. Emails: {\sffamily snassirpour@sdsu.edu}, {\sffamily tnguyen58@sdsu.edu}, and {\sffamily duy.nguyen@sdsu.edu}.}

\thanks{This work was supported in part by the U.S. Army Research Office under grant W911NF2310226, and in part by the U.S. National Science Foundation under grants ECCS-2146436, CCF-2225576, and CCF-2322190. Part of this work was presented at the IEEE International Conference on Communications (ICC) Workshops, Denver, CO, USA, June 2024 \cite{VB_Time_Varying_2024_ICC_Wrkshp}.}

}
}
	
	\maketitle
	
	\begin{abstract} 
        \subfile{Sections/Abstract}
	\end{abstract}
	
	\section{Introduction}
        \label{Sec:intro}
        \subfile{Sections/Introduction}

	
	\section{System Model}
        \label{Sec:system_model}
        \subfile{Sections/System_Model}
 
	
	\section{Online Processing Strategy}
        \label{Sec:VB_for_JED}
        \subfile{Sections/VB_for_JED}
    \section{Block Processing Strategy}
        \label{Sec:VB_for_JED_BP}
        \subfile{Sections/VB_for_JED_Block_Processing}
 	\section{Simulation Results}
        \label{Sec:numerical}
        \subfile{Sections/Numerical_Results}
        
 	\section{Conclusion}
        \label{Sec:conclusion}
        \subfile{Sections/Conclusion}
        
        \section*{Acknowledgment}
        The authors would like to thank the anonymous reviewers for their insightful and constructive comments, which helped to improve the 
 content and presentation of the paper. 

	\bibliographystyle{IEEEtran}
        \bibliography{refs}

\end{document}

%% file: Sections/Abstract.tex
Massive multiple-input multiple-output (MIMO) stands as a key technology for advancing performance metrics such as data rate, reliability, and spectrum efficiency in the fifth generation (5G) and beyond of wireless networks. However, its efficiency depends greatly on obtaining accurate channel state information (CSI). This task becomes particularly challenging with increasing user mobility. In this paper, we focus on an uplink scenario in which a massive MIMO base station serves multiple high-mobility users.  
We leverage variational Bayesian (VB) inference for joint channel estimation and data detection (JED), tailored for time-varying channels. In particular, we use the VB framework to provide approximations of the true posterior distributions. To cover more real-world scenarios, we assume the time correlation coefficients associated with the channels are unknown. Our simulations demonstrate the efficacy of our proposed VB-based approach in tracking these unknown time correlation coefficients. We present two processing strategies within the VB framework: online and block processing strategies. The online strategy offers a low-complexity solution for a given time slot, requiring only the knowledge of the parameters/statistics within that time slot. In contrast, the block processing strategy focuses on the entire communication block and processes all received signals together to reduce channel estimation errors. Additionally, we introduce an interleaved structure for the online processing strategy to further enhance its performance. Finally, we conduct a comparative analysis of our VB approach against the linear minimum mean squared error (LMMSE), the Kalman Filter (KF), and the expectation propagation (EP) methods in terms of symbol error rate (SER) and channel normalized mean squared error (NMSE). Our findings reveal that our VB framework surpasses these benchmarks across the performance metrics.

%% file: Sections/Introduction.tex
The International Telecommunication Union (ITU) estimates that by 2030, over 17 billion devices, such as mobile phones, unmanned aerial vehicles (UAVs), Internet of Things (IoT) devices, etc., will be supported via the fifth generation (5G) and beyond of wireless networks~\cite{kadir2021b5g}. Consequently, there is a pressing need to enhance key performance indicators like data rates, reliability, and spectrum efficiency in 5G networks. To meet these requirements, a widely recognized solution is massive multiple-input multiple-output (MIMO)~\cite{boccardi2014five,tan2017spectral,lu2014overview, gupta2023greenmo}. This involves deploying an array of antennas at the base station (BS), enabling the simultaneous transmission of multiple data streams to multiple users, thereby vastly increasing spectral efficiency and network capacity.

To achieve the aforementioned potential gains, accurate channel state information (CSI) is essential for both users, aiding in signal detection, and the BS, facilitating effective beamforming. This imperative underscores the need for effective algorithms for channel estimation and data detection. However, the adoption of massive MIMO results in large channel matrices, posing a significant challenge in obtaining CSI. In uplink scenarios, CSI acquisition is typically performed via known pilot signals, but this method faces scalability issues as it requires numerous orthogonal pilot signals equal to or greater than the number of users. Such constraints can lead to excessive overhead, consequently diminishing spectral efficiency as the number of users increases. 



\subsection{State-of-the-art}
To tackle the scalability challenge of channel estimation, previous studies have proposed blind channel estimation algorithms~\cite{ghavami2017blind,chen2019blind}, which rely solely on received signals. Nevertheless, these approaches are vulnerable to phase ambiguities present in the demodulated symbols. Semi-blind channel estimation offers an alternative method to mitigate the reliance on orthogonal pilot signals. This approach integrates a limited number of known pilot signals with received signals to refine the accuracy of the estimation process. In~\cite{de1997cramer}, the research presented Cramer-Rao bounds (CRBs) for channel estimation methods using semi-blind, blind, and training sequence approaches in a single-input multiple-output (SIMO) network. Next, in~\cite{de1997asymptotic}, the authors investigated two semi-blind channel estimation techniques based on maximum likelihood (ML) estimation under deterministic and Gaussian models. Thereafter, in~\cite{nayebi2017semi}, the emphasis was on semi-blind channel estimation for multi-user MIMO systems and the authors proposed two methods based on the expectation maximization (EM) framework to estimate the channels. Subsequently, in~\cite{hu2015semi}, the focus was placed on channel estimation within an uplink multi-cell massive MIMO network. The authors presented a method that first detects the data from the target cell and then computes the least square channel estimate by treating the detected data as pilot signals.  


While individual methods for channel estimation and data detection can yield promising results, joint channel estimation and data detection (JED) is often preferred in practical settings. The JED approach achieves lower channel estimation error and improved spectral efficiency compared to individual channel estimation methods. Traditional approaches rely solely on pilot signals, where longer pilot signals are desirable to reduce estimation error. However, the long sequence of known pilot signals degrades spectral efficiency. The JED approach overcomes this by generating long semi-pilot signals through the combination of short pilot signals with detected unknown data symbols. The JED approach leverages the interdependency between channels and data to continuously refine the performance of the channel estimation and data detection processes, improving overall system performance while maintaining high spectral efficiency. Thus, the research conducted in~\cite{liang2019semi} delved into the challenge of JED in hybrid massive MIMO systems. It introduced two iterative algorithms employing a low-rank matrix completion method. 
Next, the study in \cite{kuai2020double} focused on a massive MIMO system with generalized spatial modulation (GSM). It assumed the angular channel model with double sparsity in both the channels and signals, and developed two methods, blind and semi-blind, based on message passing to jointly perform user activity estimation and JED. Then, the authors in \cite{park2017expectation} studied the JED problem in an orthogonal frequency division multiplexing (OFDM) system. They considered the extended vehicular A (EVA) channel model 
and proposed a method based on the EM technique. This approach identifies reliable data tones from both desired and interfering users, which are then used as pilots for channel re-estimation.

The above algorithms assume that channels remain unchanged over the communication time, which holds true only if users are stationary or have low mobility. However, 5G networks are anticipated to support high-mobility users, e.g., UAVs and high-speed trains, causing the communication channels to fluctuate over time due to Doppler spread. Therefore, efficient channel estimation protocols are essential for promptly updating channel coefficients throughout the communication time. In~\cite{chopra2017performance, papazafeiropoulos2016impact, truong2013effects}, the authors focused on time-varying channels as a model that reflects channel aging in massive MIMO networks. They introduced various channel estimation techniques based on linear minimum mean squared error (LMMSE). Then, the work in \cite{ma2018sparse} introduced an EM-based sparse Bayesian learning (SBL) method for time-varying channels, tailored for both uplink and downlink communications. It reformulated time-varying channels as a sparse signal model using the virtual channel representation and then followed the SBL method to estimate the time-varying channels. Subsequently,~\cite{osinsky2020bayesian,lerch2015experimental} presented interpolation-based strategies for channel estimation. The effectiveness of these methods heavily depends on the number of missing channel coefficients between pilot signals as well as the length of the pilot signals. Hence, in~\cite{peng2017channel}, the authors proposed a first-order Taylor expansion channel prediction (FIT-CP) method, which utilizes the first-order Taylor expansion to predict the missing channel coefficients in the interpolation-based framework. The authors demonstrated in~\cite{peng2017channel} that their approach outperforms interpolation methods as well as traditional channel estimation approaches. Additionally,~\cite{kashyap2017performance,srivastava2018quasi} used Kalman Filter (KF) for channel estimation in time-varying channels. Later,~\cite{dong2019deep,liao2019deep} proposed deep learning-based approaches for training neural networks tailored to the dynamics of time-varying channels. Recently, \cite{kim2022semi} proposed a reinforcement learning (RL)-based data-aided channel estimator for time-varying channels by formulating the channel estimation problem as a Markov decision process (MDP) and developing an RL algorithm to optimize its policy.

In the above algorithms, channel estimation relies solely on pilot signals, while JED techniques enhance performance by leveraging both pilots and data symbols. Therefore,~\cite{naraghi2021semiICC} focused on time-varying channels in the uplink of multi-cell massive MIMO systems and proposed a semi-blind JED technique, based on the expectation propagation (EP) algorithm. 
The EP algorithm is an iterative approach that uses the factorization structure of the true distributions in order to approximate the joint a-posteriori distribution of the unknown channel matrix and data symbols~\cite{minka2013expectation,qi2007window}. The authors in~\cite{naraghi2021semiICC} assumed that the time correlation coefficients associated with the channels are known and fixed. However, in practice, these coefficients could be unknown and may even vary slowly during the communication time. As a result, in this paper, we study an uplink scenario in massive MIMO systems under time-varying channels with unknown time correlation coefficients. We introduce a technique based on variational Bayesian (VB) inference, enabling JED for time-varying channels. The VB method serves as a statistical inference framework, tackling the challenge of approximating the posterior distribution of latent variables by optimizing simpler distributions from a known family to replace the intractable true posterior distributions. VB inference, originating from machine learning, has also gained attention in wireless communications~\cite{thoota2021variational,nguyen2022variational, wan2022joint, zhu2019grid, bera2023iterative, liu2020uplink}. The research in \cite{wan2022joint} applied the VB method for performing joint carrier frequency offset (CFO) estimation and JED in an underwater acoustic OFDM system. Then, in \cite{zhu2019grid}, the authors introduced a quantized version of VB to tackle the channel estimation problem in millimeter-wave (mmWave) MIMO systems, where each antenna element is equipped with a low-resolution analog-to-digital converter (ADC). Furthermore, in \cite{bera2023iterative}, the authors studied two-way relaying time-invariant mmWave channels and proposed a VB-based approach for channel estimation. Additionally, the study in \cite{liu2020uplink} investigated massive MIMO systems with orthogonal time-frequency space (OTFS) modulation and adopted an EM-based VB framework to estimate uplink sparse channel parameters.

In this manuscript, we leverage the VB framework to address the JED problem in time-varying channels and show via simulations that our VB-based method outperforms state-of-the-art benchmarks.

\subsection{Contribution} We provide more details about our contributions below:

\begin{itemize}
    \item We devise a JED technique using VB inference for uplink scenarios in massive MIMO systems, suited for high-mobility users experiencing time-varying channels.
    \item We consider that the BS does not possess knowledge of the noise variance, emphasizing the impact of residual inter-user interference in our calculations. Additionally, to better reflect real-world conditions, we assume the time correlation coefficients corresponding to the communication channels are unknown and let the VB framework estimate them. Our simulations demonstrate that the proposed VB method yields nearly identical performance in terms of symbol error rate (SER) compared to a scenario where the time correlation coefficients are known.
    \item As the computational complexity could lead to a delay in the channel estimation process, and thus the risk of outdated CSI, we design an online VB processing strategy using only the parameters/statistics of each time slot. We also develop an interleaved structure to further improve its performance.
    \item Since CSI accuracy is essential in massive MIMO systems, we offer a block processing strategy based on VB inference, which analyzes all received signals together to enhance the channel estimation performance.
    \item We compare the performance of our proposed VB method with the LMMSE, KF, and EP methods in terms of SER and channel normalized mean squared error (NMSE). We show that our VB method outperforms the benchmarks in all performance metrics.
\end{itemize}

The rest of this paper is organized as follows. In Section~\ref{Sec:system_model}, we provide the system model. Then, in Sections~\ref{Sec:VB_for_JED} and~\ref{Sec:VB_for_JED_BP}, we explain the details of our proposed online and block processing strategies. Section~\ref{Sec:numerical} includes the simulation results of this work, and Section~\ref{Sec:conclusion} concludes this paper.

\subsection{Notation} In this paper, we use italic, bold lowercase, and bold uppercase font styles to signify scalars, vectors, and matrices, respectively. We utilize the notation $\mathcal{CN}(\bs{\mu},\bs{\Sigma})$ to denote a complex Gaussian random vector with a mean vector $\bs{\mu}$ and a covariance matrix $\bs{\Sigma}$. $\Gamma(a, b)$ implies a Gamma distribution parametrized by $a$ and $b$. The space of $x\times y$ dimensional complex-valued matrices is denoted by $\mathbb{C}^{x\times y}$. We use $\sim$ and $\propto$ to convey the notations of ``distributed according to'' and ``proportional to,'' respectively. $\mathrm{diag}\{\textbf{a}\}$ is a diagonal matrix based on $\textbf{a}$. $|\textbf{A}|$, $\|\textbf{A}\|$, $\|\textbf{A}\|_{\mr{F}}$, $\textbf{A}^{\top}$, and $\textbf{A}^H$ represent the determinant, Euclidean norm, Frobenius norm, transpose, and conjugate transpose of $\textbf{A}$, respectively. The notation $[\mb{X}]_{ij}$ denotes the element located at the $i^{\mathrm{th}}$ row and $j^{\mathrm{th}}$ column of $\mb{X}$. $\mr{Tr}\left(\cdot\right)$ is the trace function, $\mr{exp}\left\{\cdot\right\}$ denotes the exponential function, $\log_{10}\left(\cdot\right)$ is the logarithmic function with base $10$, and $\textbf{I}_M$ stands for the identity matrix of size $M$. We use $x^{\star}$ and $\Re \{x\}$ to indicate the complex conjugate and the real part of $x$, respectively. The probability density function (PDF) of a length-$K$ complex-valued random vector $\mathbf{x} \sim\mathcal{CN}(\bs{\mu},\bs{\Sigma})$ is given by: $\mathcal{CN}(\mb{x};\bs{\mu},\bs{\Sigma}) = \frac{1}{\pi^K|\bs{\Sigma}|}\mathrm{exp}\big(-(\mb{x}-\bs{\mu})^H\bs{\Sigma}^{-1}(\mb{x}-\bs{\mu})\big)$. $\mathbb{E}_{p(x)}[x]$ and $\mathrm{Var}_{p(x)}[x]$ represent the mean and variance of $x$ according to its distribution $p(x)$. Similarly, $\langle x\rangle$, $\langle |x|^2\rangle$, and $\tau^x$ represent the mean, second moment, and variance of $x$ based on a variational distribution $q(x)$. The subscript $t|t-1$ denotes predicted statistics at time $t$ utilizing data from time $t-1$, whereas $t|t$ denotes the posterior estimated statistics at time $t$ derived from observations made at time $t$.

%% file: Sections/System_Model.tex
\begin{figure}[t]
\centering
  \includegraphics[trim = 0mm 0mm 0mm 0mm, clip, scale=7, width=0.8\linewidth, draft=false]{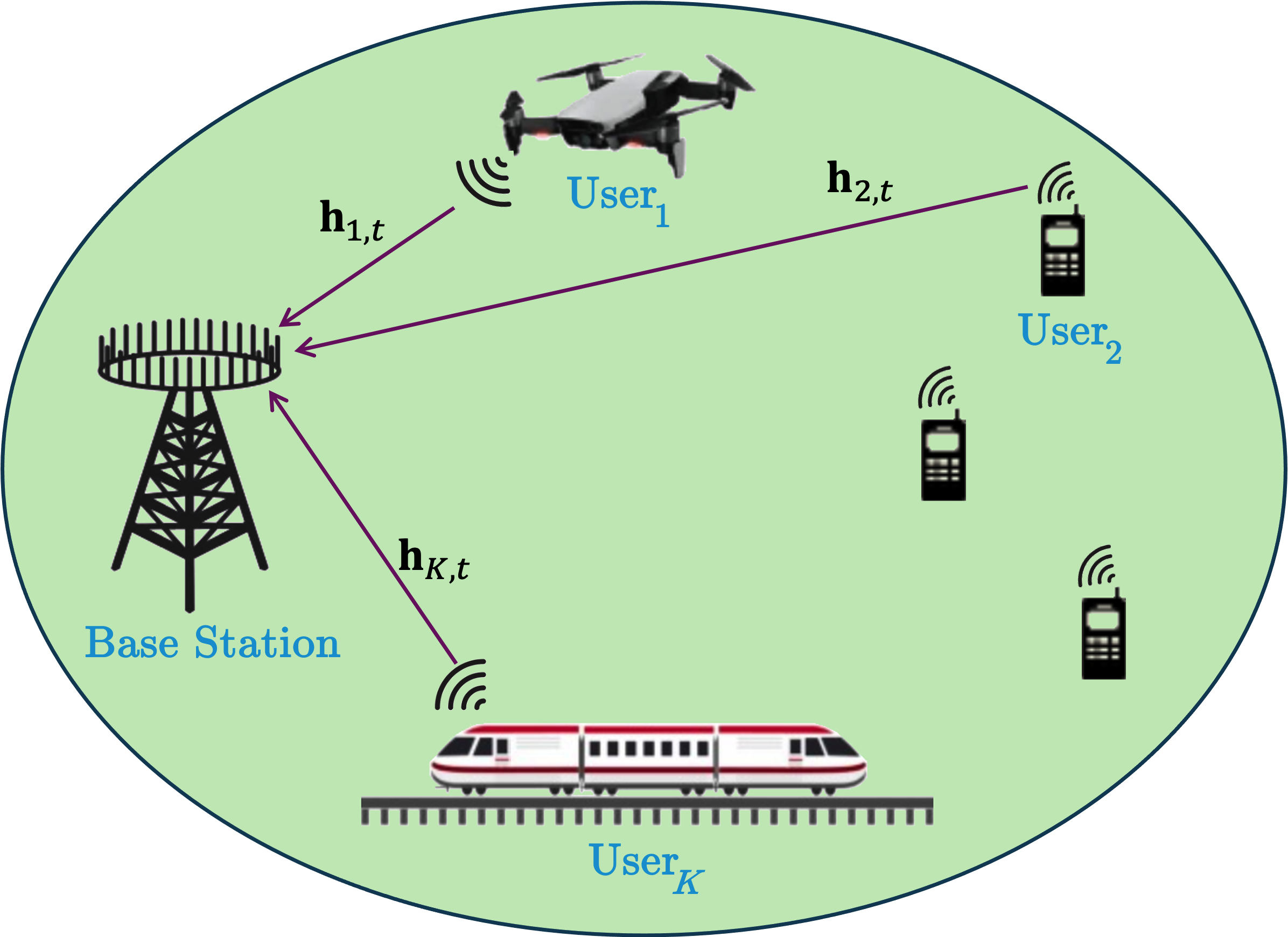}
   \vspace{-2mm}
  \caption{The uplink scenario in a massive MIMO network supporting $K$ single-antenna users, which could be high-mobility users.
  }\label{Fig:system_model}
  \vspace{-5mm}
\end{figure}

In this work, we focus on an uplink scenario in a massive MIMO network, as illustrated in Fig.~\ref{Fig:system_model}, where a BS with multiple antennas aims to communicate with multiple users simultaneously. Specifically, we address a JED problem within this network, which supports high-mobility users such as UAVs and high-speed trains that experience time-varying channels. To tackle the JED problem, we propose a method based on VB inference to approximate the true posterior distributions. We begin this section by explaining the channel model, followed by a quick introduction to VB inference, and then proceed to formulate the problem.

\subsection{Channel Model}
We use $\mb{h}_{i,t} = [h_{i,t}^{[1]}, h_{i,t}^{[2]}, \ldots, h_{i,t}^{[M]}]^{\top} \in \mathbb{C}^{M \times 1}$ to denote the communication channel between the $i^{\mathrm{th}}$ user and the BS at time $t$, where $M$ is the number of antenna elements at the BS, $i \in \{1, 2, \ldots, K\}$, and $K$ is the number of users. We assume each user has a single antenna. In this work, we utilize the first-order Gauss-Markov model to characterize $\mb{h}_{i,t}$ as follows~\cite{chopra2016throughput}: 
\begin{eqnarray} \label{GM-model}
    \mb{h}_{i,t}&=& \eta_i\mb{h}_{i,t-1} + \sqrt{1-\eta_i^2}{\mb{R}_i^{\frac{1}{2}}}\mb{g}_{i,t},~~ t=1,\ldots,T,
\end{eqnarray}
where $\mb{h}_{i,0}  = \mb{R}_i^{\frac{1}{2}}\mb{g}_{i,0}$,   $T$ is the communication time, and $\mb{g}_{i,t}$ models small-scale fading, which is independent of $\mb{h}_{i,t-1}$ and assumed to be Rayleigh fading (i.e., $\mb{g}_{i,t} \sim \mathcal{CN}(\mb{0}, \mb{I}_M)$). Further, $\eta_i$ denotes the time correlation coefficient of the channels corresponding to the $i^{\mathrm{th}}$ user, which is given by~\cite{chopra2016throughput}:
\begin{align}
    \eta_i = J_0(2\pi f^d_i T_s),
\end{align}
in which $J_0(\cdot)$ denotes the zero-order Bessel function
of the first kind, $f^d_i$ is the maximum Doppler frequency corresponding to the channels of the $i^{\mathrm{th}}$ user, and $T_s$ is the sampling period. Moreover, in~\eqref{GM-model}, $\mb{R}_i$ represents the covariance of $\mb{h}_{i,t}$ (i.e., $\mathbb{E}\left[\mb{h}_{i,t}\mb{h}_{i,t}^H\right] = \mb{R}_i$), which is equal to:
\begin{align}
    \mb{R}_i = \beta_i^{-1}\mb{R},
\end{align}
where $\beta_i$ is the large-scale fading for the propagation channel from the $i^{\mathrm{th}}$ user, and $\mb{R}$ denotes the spatial correlation at the BS. 
We assume that $\mb{R}_i$ is known at the BS and use it to compute the autocovariance between $\mb{h}_{i,t-1}$ and $\mb{h}_{i,t}$, which follows $\mathbb{E}\left[\mb{h}_{i,t-1}\mb{h}_{i,t}^H\right] = \eta_i\mb{R}_i$. Then, we have:
\begin{align}
\label{eq_cond_ht_ht_1}
    &p(\mb{h}_{i,t}|\mb{h}_{i,t-1},\eta_i;\mb{R}_i) = \mc{CN}\left(\mb{h}_{i,t};\eta_i\mb{h}_{i,t-1},  (1-\eta_i^2)\mb{R}_i\right).
    \end{align} 
    
We also assume a channel independence among users, implying that $\mathbb{E}[\mb{h}_i\mb{h}_j^H] = \mb{0}$, $1 \leq i,j \leq K$ and $i\neq j$.

{\bf Received signal:} In our massive MIMO network, the BS captures $y_{m,t}, 1 \leq m \leq M,$ as the received signal through its $m^{\mathrm{th}}$ antenna element at time $t$, which can be modeled as:
\begin{eqnarray}
\label{system-model_signal_m}
    y_{m,t} = \sum_{i=1}^K h_{i,t}^{[m]} x_{i,t} + n_{m,t},
\end{eqnarray}
where $h_{i,t}^{[m]}$ is the $m^{\mathrm{th}}$ element of $\mb{h}_{i,t}$, $x_{i,t}$ is the transmitted signal from the $i^{\mathrm{th}}$ user, and $n_{m,t} \sim \mathcal{CN}(0, N_0)$ denotes the independent and identically distributed (i.i.d.) additive white Gaussian noise (AWGN) at the $m^{\mathrm{th}}$ antenna element during time $t$. Subsequently, we arrange the received signals, transmitted signals, communication channels, and noise components into a vector format to establish a linear model for the received signals as below:
\begin{eqnarray}
\label{system-model}
    \mb{y}_t = \mb{H}_t\mb{x}_t + \mb{n}_t,
\end{eqnarray}
where $\mb{y}_t = [y_{1,t}, y_{2,t}, \ldots y_{M,t}]^{\top} \in\mathbb{C}^{M\times 1 }$, $\mb{x}_t=[x_{1,t},x_{2,t},\ldots,x_{K,t}]^{\top}\in \mathbb{C}^{K\times 1 }$, $\mb{H}_t=\left[\mb{h}_{1,t},\mb{h}_{2,t},\ldots,\mb{h}_{K,t}\right]\in \mathbb{C}^{M\times K }$, and $\mb{n}_t = [n_{1,t}, n_{2,t}, \ldots n_{M,t}]^{\top}\in\mathbb{C}^{M\times 1 }$. 

In this paper, we concentrate on time-varying channels with the aim of jointly estimating $\mb{H}_t$ and detecting $\mb{x}_t$ using the received signal $\mb{y}_t$. To accomplish this, we propose a method based on VB inference. Before delving into the details, we provide a brief overview of VB.

\subsection{An Overview of VB inference} 
Variational Bayesian (VB) inference stands as a robust statistical framework derived from machine learning, devised to address the challenge of approximating the posterior distribution of latent variables. Here, we denote the set of all observed variables as $\mb{y}$ and the set of $V$ latent variables as $\mb{x}$. To detect $\mb{x}$, it is crucial to determine the posterior $p(\mb{x}|\mb{y})$, which often proves computationally intractable. To tackle this challenge, the VB method endeavors to identify a distribution $q(\mb{x})$ characterized by variational parameters within a predefined family $\mc{Q}$ of densities, such that $q(\mb{x})$ closely approximates $p(\mb{x}|\mb{y})$. To this end, VB defines an optimization problem leveraging the Kullback-Leibler (KL) divergence from $q(\mb{x})$ to $p(\mb{x}|\mb{y})$:
\begin{equation}
\label{eq:opt_KL}
q^{\star}(\mb{x}) = \arg\min_{q(\mb{x}) \in \mc{Q}} \text{KL}\big(q(\mb{x}) \| p(\mb{x}|\mb{y}) \big),
\end{equation}
where $q^{\star}(\mb{x})$ denotes the optimal variational distribution, and
\begin{equation}
\text{KL}\big(q(\mb{x}) \| p(\mb{x}|\mb{y}) \big) = \mathbb{E}_{q(\mb{x})} \big[\ln q(\mb{x}) \big] - \mathbb{E}_{q(\mb{x})} \big[\ln p(\mb{x}|\mb{y})\big].
\end{equation}

The minimization of \eqref{eq:opt_KL} occurs when $q(\mb{x}) = p(\mb{x}|\mb{y})$. However, because obtaining the true posterior distribution is often infeasible, it is more practical to employ a restricted family of distributions $q(\mb{x})$. Thus, the literature~\cite{Bishop-2006} commonly focuses on the mean-field variational family, defined as:
\begin{equation}
q(\mb{x}) = \prod_{i=1}^V q_i(x_i),
\end{equation}
where the latent variables are assumed to be mutually independent, each governed by a distinct factor in the variational density. In this case, the optimal value of $q_i(x_i)$ is equal to~[31, Chapter 10]:
\begin{align}
 \label{eq:q_start_prop}
		q_i^{\star}(x_i) &\propto \mr{exp}\left\{\big\langle{\ln p (\mb{y},\mb{x})\big\rangle}\right\},
\end{align}
where $\lr{\cdot}$ is the expectation across all latent variables excluding $x_i$, utilizing the currently fixed variational density $q_{-i}(\mb{x}_{-i}) = \prod_{j=1, j\neq i}^{V} q_{j}(x_{j})$. Then, to optimize~\eqref{eq:opt_KL}, we use the Coordinate Ascent Variational Inference (CAVI) algorithm, which is an iterative method that ensures convergence to at least a locally optimal solution~\cite{Wainwright-2008}. In particular, the CAVI algorithm updates $q_i^{\star}(x_i)$ sequentially for all latent variables to monotonically enhance the objective function in~\eqref{eq:opt_KL}.


\subsection{Problem Formulation}
According to \eqref{eq:q_start_prop}, in order to apply the VB framework, it is necessary to compute the joint distribution. Unlike previous studies~\cite{naraghi2021semi, thoota2021variational}, we assume that the time correlation $\eta_i$ and the noise variance $N_0$ are i.i.d. random variables and unknown a-priori at the BS. We denote the precision of the noise at time $t$ as $\gamma_t = 1/N_0$. Utilizing these assumptions, we derive the joint distribution $p(\mb{y}_t, \mb{H}_t, \mb{H}_{t-1}, \mb{x}_t, \gamma_t, \bs{\eta}; \bar{\mb{R}})$ as follows:
\begin{align}
\label{eq:first_cond_prob}	&p(\mb{y}_t,\mb{x}_t,\mb{H}_t,\mb{H}_{t-1},\gamma_t,\bs{\eta};\bar{\mb{R}}) = p(\mb{y}_t|\mb{x}_t,\mb{H}_t,\gamma_t)p(\mb{x}_t) \\
    &\quad \quad \quad \quad \quad ~~\times \underbrace{p(\mb{H}_{t}|\mb{H}_{t-1}) p(\mb{H}_{t-1}|\bs{\eta};\bar{\mb{R}})}_{p(\mb{H}_{t}|\bs{\eta};\bar{\mb{R}})}p(\gamma_t)p(\bs{\eta}),\nonumber
\end{align}
where $p(\mb{H}_{t}|\bs{\eta};\bar{\mb{R}}) = \prod_{i=1}^K p(\mb{h}_{i,t}| \mb{h}_{i,t-1})p(\mb{h}_{i,t-1}|\eta_i;\mb{R}_i)$, $\bs{\eta}=[\eta_1, \eta_2,\ldots,\eta_K]$, $\bar{\mb{R}}=[\mb{R}_1, \mb{R}_2,\ldots,\mb{R}_K]$, $p(\mb{x}_t) = \prod_{i=1}^{K}p(x_{i,t})$, and $p(\bs{\eta}) = \prod_{i=1}^{K}p(\eta_i)$. 

In this paper, we utilize the
VB framework to approximate the posterior
distribution $p(\mathbf{x}_t,\mathbf{H}_t, \bs{\eta},\gamma_t|\mathbf{y}_t)$. There are two possible strategies for processing time-varying channels: one that utilizes information available at time $t$, which we refer to as the online processing strategy, and another that processes all observed signals $\mathbf{y}_t$'s together, which we call the block processing strategy. Each of these strategies has its own advantages and drawbacks. We explain the details of these strategies in the next two sections.
 

%% file: Sections/VB_for_JED.tex
In this section, we explain the details of our proposed VB-based online processing strategy designed to address the JED problem in time-varying channels. The primary objective of the online processing strategy is to achieve a low-complexity low-latency solution. To do so, we propose a method that only requires the parameters and statistics corresponding to the given time $t$. Since estimating the channel and detecting data based on the limited statistics available at time $t$ is challenging, we introduce a two-phase online processing strategy consisting of a prediction phase and an estimation phase.

The prediction phase aims to provide a dynamic prior distribution for $p(\mathbf{h}_{i,t} | \eta_i, \mathbf{R}_i)$ instead of assuming a fixed prior distribution. To achieve this, we continuously update the prior distribution for $(\mathbf{h}_{i,t} | \eta_i, \mathbf{R}_i)$ using the information available at time $t-1$. Following the prediction phase, the estimation phase is employed to complete the online processing strategy by estimating the corresponding channel $\mathbf{H}_t$ and detecting the data symbol $\mathbf{x}_t$ at a given time $t$. The details of these two phases are presented below.


\textbf{Phase I - Prediction:} In \eqref{eq:first_cond_prob}, we need to find the probability 
$p(\mb{h}_{i,t}|\eta_i; \mb{R}_i) = p(\mb{h}_{i,t}|\mb{h}_{i,t-1}) (\mb{h}_{i,t-1}|\eta_i; \mb{R}_i)$ in order to compute the desired joint distribution. In this phase, our goal is to find a predictive distribution for 
 $p(\mb{h}_{i,t}|\eta_i; \mb{R}_i)$ at time 
$t$, using the statistics available at time 
$t-1$, which will later be used in Phase II. To this end, we calculate the posterior distribution of 
$p(\mb{h}_{i,t-1}|\eta_i; \mb{R}_i)$ at time slot $t-1$ using the Gauss-Markov model in \eqref{GM-model}, and then consider it as the prior distribution for 
$p(\mb{h}_{i,t}|\eta_i; \mb{R}_i)$ at time slot 
$t$. Moreover, according to the VB framework, the prior distributions of the channels are necessary to obtain their posterior distributions. In this case, we assume the prior distribution $p(\mb{h}_{i,t-1}) = \mc{CN}\big(\mb{h}_{i,t-1}; \hat{\mb{h}}_{i,t-1|t-1},\hat{\bs{\Sigma}}_{i,t-1|t-1}\big)$. If $\eta_i$ is known, we use the prior distribution $p(\mb{h}_{i,t-1})$ and the Gauss-Markov model in~\eqref{GM-model} to derive the predictive distribution $p(\mb{h}_{i,t} 
|\eta_i, \mb{R}_i)$ as below:
\begin{align}
    p(\mb{h}_{i,t} 
|\eta_i, \mb{R}_i) = \mc{CN}(\mb{h}_{i,t};\hat{\mb{h}}_{i,t|t-1},\hat{\bs{\Sigma}}_{i,t|t-1}),
\end{align}
where 
	\begin{eqnarray} 
		\hat{\mb{h}}_{i,t|t-1}&=&\eta_i\hat{\mb{h}}_{i,t-1|t-1}. \label{h-prior-mean} \\
		\hat{\bs{\Sigma}}_{i,t|t-1} &=& \eta_i^2\hat{\bs{\Sigma}}_{i,t-1|t-1} + (1-\eta_i^2)\mb{R}_i.
		\label{h-prior-cov} 
	\end{eqnarray}
 
\begin{remark}
In this work, we assume $\eta_i$ is an unknown random variable, and the prior distribution $p(\eta_i)$ is a conjugate probability. To achieve this, we assume $p(\eta_i) = \mc{N}\big(\eta_i; \hat{\eta}_{i,t-1}, \tau^{\eta}_{i,t-1}\big)$. However, this assumption may result in $\eta_{i}$ being less than $0$ or exceeding $1$, even though it must always lie within the range $0 \leq \eta_{i} \leq 1$. On the other hand, after a few iterations of the CAVI algorithm, we expect the variance $\tau^{\eta}_{i,t-1}$ to decrease progressively, causing the algorithm to concentrate around the mean $\hat{\eta}_{i,t-1}$. Therefore, it is crucial to select the initial value of $\hat{\eta}_{i,t-1}$ carefully, ensuring it is close to the true value. As a result, we reset the initial value of $\hat{\eta}_{i,t-1}$ if it is less than $0$ or exceeds $1$, which ensures that $\eta_{i}$ always remains within its acceptable range.
\end{remark}

In statistical models, when it is challenging to obtain an accurate closed-form distribution of a random variable, that variable is typically approximated by its mean value. This is based on the expectation that its variance will become progressively smaller throughout the statistical process. Therefore, in this paper, we update \eqref{h-prior-cov} by approximating the predictive covariance $\hat{\bs{\Sigma}}_{i,t|t-1}$ using the second moment of $\eta_i$ at time slot $t-1$.  In particular, we replace $\eta^2_i$ with $\mathbb{E}(\eta^2_i) = \hat{\eta}_{i,t-1}^2 + \tau^{\eta}_{i,t-1}$. Here, we have:
	\begin{align} 
	\hat{\bs{\Sigma}}_{i,t|t-1} &\approx(\hat{\eta}_{i,t-1}^2 + \tau^{\eta}_{i,t-1}) \hat{\bs{\Sigma}}_{i,t-1|t-1} \nonumber \\
   & \quad + (1- \hat{\eta}_{i,t-1}^2 - \tau^{\eta}_{i,t-1})\mb{R}_i.
		\label{h-prior-cov-2} 
	\end{align}

\textbf{Phase II - Estimation:} In this phase, our objective is to derive Bayesian optimal estimates for both the data symbol $\mathbf{x}_t$ and the channel matrix $\mathbf{H}_t$. To accomplish this, we require the posterior distribution $p(\mathbf{x}_t,\mathbf{H}_t, \bs{\eta},\gamma_t|\mathbf{y}_t)$, which could be challenging to obtain. Consequently, we employ the mean-field variational distribution $q(\mathbf{x}_t,\mathbf{H}_t, \bs{\eta},\gamma_t)$ proposed within the VB framework in the following manner:
\begin{align} \label{eq:mean_field}
p\big(\mb{x}_t,\mb{H}_t, \bs{\eta},\gamma_t|\mb{y}_t\big)
    &\approx q(\mb{x}_t,\mb{H}_t, \bs{\eta},\gamma_t) \\
    &= \left[ \prod_{i=1}^K q_i(x_{i,t}) 
    q(\mb{h}_{i,t}) q(\eta_i)\right] q(\gamma_t).\nonumber 
\end{align} 

According to~\eqref{eq:q_start_prop}, in order to get the optimal solution of the variational densities in \eqref{eq:mean_field}, we need the joint distribution $p(\mb{y}_t, \mb{x}_t,\mb{H}_t, \bs{\eta},\gamma_t)$, which is equal to:
\begin{align} \label{conditional}
    &p(\mb{y}_t, \mb{x}_t,\mb{H}_t,\gamma_t)= p(\mb{y}_t| \mb{x}_t,\mb{H}_t, \bs{\eta},\gamma_t) \nonumber \\
    &\quad \quad \quad \quad \quad  \times \left[\prod_{i=1}^K p(x_{i,t})p(\mb{h}_{i,t}|\eta_i;\mb{R}_i)p(\eta_i)  \right]p(\gamma_t).
\end{align}

In~\eqref{conditional}, besides considering $p(\mb{h}_{i,t}|\eta_i;\mb{R}_i)$ and $p(\eta_i)$, we also need to specify the prior distributions for $\gamma_t$ and $x_{i,t}$. 
\begin{remark}
    In MIMO systems, $\gamma_t$ is typically assumed to be a known and fixed parameter \cite{kashyap2017performance,naraghi2021semiICC}. However, recent studies \cite{nguyen2022variational,wan2022joint} have considered $\gamma_t$ as an unknown parameter and applied statistical methods to estimate it. Specifically, \cite{wan2022joint} utilized the EM method to approximate $\gamma_t$, while \cite{nguyen2022variational} assumed it follows a Gamma distribution and leveraged the VB method to obtain $\gamma_t$. In this paper, motivated by these works, we assume $\gamma_t \sim \Gamma(a_0, b_0)$ to capture both the impact of noise and the estimation error in our process, and we then use the CAVI algorithm to find its local optimum solution.
\end{remark}

Assumptions about the prior distribution of $x_{i,t}$ vary depending on whether we are in the pilot transmission phase or data transmission phase. In this study, we divide the total communication time $T$ into two parts: the pilot transmission phase with duration $T_p$ followed by the data transmission phase with duration $T_d$, where $T_p + T_d = T$. If $1 \leq t \leq T_p$, the prior distribution of $x_{i,t}$ is simply $p(x_{i,t}) = \delta(x_{i,t} - \bar{x}_{i,t})$, where $\bar{x}_{i,t}$ represents the pilot data symbol from the $i^{\mathrm{th}}$ user at time $t$ and $\delta(\cdot)$ is the Dirac delta function. For $T_p < t \leq T$, the prior distribution of $x_{i,t}$ is expressed as $p(x_{i,t}) = \sum_{a\in \mc{S}} p_a\delta(x_{i,t} - a)$, where $p_a$ denotes the probability of the constellation point $a\in \mc{S}$ and $\mc{S}$ represents the signal constellation.

\begin{table} [t]
\caption{The prior distribution assumptions for the unknown random variables in the online processing VB strategy
}
\centering
{
\begin{tabular}{|c|c|}
\hline
Probability & Prior Distribution Assumption\\
\hline \hline
$p(\mb{h}_{i,t-1})$ &  $\mc{CN}\big(\mb{h}_{i,t-1}; \hat{\mb{h}}_{i,t-1|t-1},\hat{\bs{\Sigma}}_{i,t-1|t-1}\big)$\\

$p(\eta_i)$ & $\mc{N}\big(\eta_i; \hat{\eta}_{i,t-1}, \tau^{\eta}_{i,t-1}\big)$ \\
$p(\gamma_t)$ & $\Gamma(a_0,b_0)$\\

$p(x_{i,t}) (\text{pilot transmission})$ & $\delta(x_{i,t} - \bar{x}_{i,t})$\\

$p(x_{i,t}) (\text{data transmission})$ & $\sum_{a\in \mc{S}} p_a\delta(x_{i,t} - a)$\\

\hline
\end{tabular}
}
\label{table:prior_dist_online}
\end{table} 

In Table~\ref{table:prior_dist_online}, we summarize the prior distribution assumptions for the desired unknown random variables $p(\mb{h}_{t-1})$, $p(\eta_i)$, $p(\gamma_t)$, and $p(x_{i,t})$.

As previously discussed in Section~\ref{Sec:system_model}, the CAVI algorithm iteratively converges to local optimum solutions  $q^{\star}(\mb{h}_{i,t})$, $q^{\star}(\eta_i)$, $q^{\star}(x_{i,t})$, and $q^{\star}(\gamma_t)$ by optimizing one latent variable at a time while keeping the others fixed. In the subsequent parts, we illustrate the update process for each latent variable.

\subsubsection{ Updating $\mb{h}_{i,t}$} By computing the expectation of~\eqref{conditional} for all latent variables except for $\mb{h}_{i,t}$, we express the variational distribution $q(\mb{h}_{i,t})$ as~\eqref{q-h-i} (at the top of the next page), 
\begin{figure*}
    \begin{align} \label{q-h-i}
        q(\mb{h}_{i,t}) &\propto \mr{exp}\Big\{\big\langle\ln p(\mb{y}_t|\mb{x}_t,\mb{H}_t,\gamma_t) + \ln p(\mb{h}_{i,t}|\eta_i;\mb{R}_i) \big\rangle\Big\}  \nonumber \\
        &\propto \mr{exp}\Big\{- \blr{\gamma_t \big\|\mb{y}_t-\mb{H}_t\mb{x}_t\big\|^2} -  \bblr{ (\mb{h}_{i,t}-\eta_i\hat{\mb{h}}_{i,t-1|t-1})^H\hat{\bs{\Sigma}}^{-1}_{i,t|t-1}(\mb{h}_{i,t}-\eta_i\hat{\mb{h}}_{i,t-1|t-1})}  \Big\} \nonumber \\
        &\propto \mr{exp}\bigg\{- \bblr{\gamma_t \bigg\|\mb{y}_t - \mb{h}_{i,t}x_{i,t} - \sum_{j\neq i}^K \mb{h}_{j,t}x_{j,t}\bigg\|^2} - \bblr{(\mb{h}_{i,t}-\eta_i\hat{\mb{h}}_{i,t-1|t-1})^H\hat{\bs{\Sigma}}^{-1}_{i,t|t-1}(\mb{h}_{i,t}-\eta_i\hat{\mb{h}}_{i,t-1|t-1})} \bigg\} \nonumber \\
        &\propto \mr{exp}\bigg\{-\mb{h}_{i,t}^H\Big[\lr{\gamma_t}\lr{|x_{i,t}|^2}\mb{I}_M+ \hat{\bs{\Sigma}}^{-1}_{i,t|t-1}\Big]\mb{h}_{i,t}  \nonumber \\
        &\quad\quad + 2\,\Re\bigg\{\lr{\gamma_t}\mb{h}_{i,t}^H\bigg(\mb{y}_t- \sum_{j\neq i}^K \lr{\mb{h}_{j,t}}\lr{x_{j,t}}\bigg)\lr{x_{i,t}^*} + \lr{\eta_i}\mb{h}_{i,t}^H\hat{\bs{\Sigma}}^{-1}_{i,t|t-1}\hat{\mb{h}}_{i,t-1|t-1}\bigg\}\bigg\}.
    \end{align}
    \hrulefill
    \vspace{-3mm}
\end{figure*}
which shows that $q(\mb{h}_{i,t})$ is Gaussian with the subsequent covariance matrix and mean: 
\begin{align}
    \bs{\Sigma}_{i,t}  &=\Big[\lr{\gamma_t}\lr{|x_{i,t}|^2}\mb{I}_M + \hat{\bs{\Sigma}}^{-1}_{i,t|t-1}\Big]^{-1},  \label{Sigma-h} \\
    \lr{\mb{h}_{i,t}} &=\bs{\Sigma}_{i,t}\Bigg[\lr{\gamma_t}\bigg(\mb{y}_t- \sum_{j\neq i}^K \lr{\mb{h}_{j,t}}\lr{x_{j,t}}\bigg)\lr{x_{i,t}^*} \nonumber \\
    &\quad\quad\quad\quad + \lr{\eta_i} \hat{\bs{\Sigma}}^{-1}_{i,t|t-1}\hat{\mb{h}}_{i,t-1|t-1}\Bigg].  \label{mean-h}
\end{align}

\subsubsection{ Updating $\eta_i$} If we take the expectation of~\eqref{conditional} with respect to all latent variables except for $\eta_i$, we obtain the variational distribution $q(\eta_i)$ as~\eqref{eta-i} (on the next page), 
\begin{figure*}
		\begin{eqnarray} 
  \label{eta-i}
			q(\eta_i) &\propto& \mr{exp}\Big\{\blr{ \ln p\big(\mb{h}_{i,t}|\eta_i;\mb{R}_i\big) + \ln p\big(\eta_i\big)}\Big\}  \nonumber \\
			&\propto& \mr{exp}\Big\{ - \Blr{\big(\mb{h}_{i,t}-\eta_i\hat{\mb{h}}_{i,t-1|t-1}\big)^H\hat{\bs{\Sigma}}_{i,t|t-1}^{-1}\big(\mb{h}_{i,t}-\eta_i\hat{\mb{h}}_{i,t-1|t-1}\big)}  - \|\eta_i-\hat{\eta}_{i,t-1}\|^2 / \tau^{\eta}_{i,t-1}\Big\} \nonumber \\
			&\propto& \mr{exp}\bigg\{ - \eta_i^2\Big(\hat{\mb{h}}_{i,t-1|t-1}^H\hat{\bs{\Sigma}}_{i,t|t-1}^{-1}\hat{\mb{h}}_{i,t-1|t-1} + 1/\tau^{\eta}_{i,t-1}\Big) +  2\eta_i\bigg(\Re\big\{\hat{\mb{h}}_{i,t-1|t-1}^H\hat{\bs{\Sigma}}_{i,t|t-1}^{-1}\lr{\mb{h}_{i,t}} \big\} + \frac{\hat{\eta}_{i,t-1}}{\tau^{\eta}_{i,t-1}}\bigg) \bigg\},
		\end{eqnarray}
  \hrulefill
  \vspace{-3mm}
  \end{figure*}
which is Gaussian. In this case, the variance and mean of $q(\eta_i)$ are given by:
\begin{align}
\label{tau_i_var}
    \tau^{\eta}_{i,t} = \left(\hat{\mb{h}}_{i,t-1|t-1}^H\hat{\bs{\Sigma}}^{-1}_{i,t|t-1}\hat{\mb{h}}_{i,t-1|t-1} + 1/\tau^{\eta}_{i,t-1}\right)^{-1},
\end{align}
\begin{align}
\label{tau_i_mean}
    \lr{\eta_i} = \tau^{\eta}_{i,t}  \left(\Re\Big\{\hat{\mb{h}}_{i,t-1|t-1}^H\hat{\bs{\Sigma}}^{-1}_{i,t|t-1}\lr{\mb{h}_{i,t}}  \Big\}+ \frac{\hat{\eta}_{i,t-1}}{\tau^{\eta}_{i,t-1} }\right).
\end{align}


We utilize the following lemma to compute the variational posterior mean of several random variables. We employ this lemma later in this section to update $x_{i,t}$ and $\gamma_t$. 
\begin{lemma}
\label{theorem-2}
\cite{nguyen2022variational} Let $\mb{A}\in \mathbb{C}^{m\times n}$ and $\mb{x}\in \mathbb{C}^{n\times 1}$ be two independent random matrices (vectors) with respect to a variational density $q_{\mb{A},\mb{x}}(\mb{A},\mb{x}) = q_{\mb{A}}(\mb{A})q_{\mb{x}}(\mb{x})$. Suppose $\mb{A}$ is column-wise independent and $\lr{\mb{a}_i}$ and $\hat{\bs{\Sigma}}_{\mb{a}_i}$ are the variational mean and covariance matrix of the $i^{th}$ column of $\mb{A}$. Let $\lr{\mb{x}}$ and $\hat{\bs{\Sigma}}_{\mb{x}}$ be the variational mean and covariance matrix of $\mb{x}$. Consider $\mb{y}\in \mathbb{C}^{m\times 1}$ is an arbitrary vector. Here, $\blr{\|\mb{y}-\mb{A}\mb{x}\|^2}$,  the expectation of $\|\mb{y}-\mb{A}\mb{x}\|^2$ with respect to $q_{\mb{A},\mb{x}}(\mb{A},\mb{x})$, is given by:
\begin{align}\label{f-ABC}
    \blr{\|\mb{y}-\mb{A}\mb{x}\|^2} &= \|\mb{y}-\lr{\mb{A}}\lr{\mb{x}}\|^2 + \lr{\mb{x}}^H\mb{D}\lr{\mb{x}}   \nonumber \\
    &\quad+\tr\big\{\hat{\bs{\Sigma}}_{\mb{x}}\mb{D}\big\}+ \tr\big\{\hat{\bs{\Sigma}}_{\mb{x}}\lr{\mb{A}^H}\lr{\mb{A}} \big\},
\end{align}
where $\mb{D} = \mr{diag}\big(\tr\{\bs{\Sigma_{\mb{a}_1}}\},\ldots, \tr\{\bs{\Sigma_{\mb{a}_n}}\}\big)$. 
\end{lemma}
\begin{IEEEproof}
We first expand $\blr{\|\mb{y}-\mb{A}\mb{x}\|^2}$ as follows: 
\begin{align}\label{expand}
    \blr{\|\mb{y}-\mb{A}\mb{x}\|^2} &= \|\mb{y}\|^2 - 2\,\Re\big\{\mb{y}^H\lr{\mb{Ax}}\big\} + \lr{\mb{x}^H\mb{A}^H\mb{A}\mb{x}} \nonumber \\ 
    &= \|\mb{y}-\lr{\mb{A}}\lr{\mb{x}}\|^2 -  \tr\big\{\lr{\mb{A}^H}\lr{\mb{A}}\lr{\mb{x}}\lr{\mb{x}^H}\big\} \nonumber \\ 
    & +\tr\big\{\lr{\mb{A}^H\mb{A}}\lr{\mb{x}\mb{x}^H}\big\}.
\end{align}

We know that $\lr{\mb{x}\mb{x}^H} = \lr{\mb{x}}\lr{\mb{x}^H} + \hat{\bs{\Sigma}}_{\mb{x}}$. Moreover, we have:
\begin{eqnarray}
\label{eq:eq_num2}
    [\lr{\mb{A}^H\mb{A}}]_{ij} &=& \left\{ \begin{array}{ll}
        \lr{\|\mb{a}_i\|^2}, &  i = j \\
        \lr{\mb{a}_i^H\mb{a}_j}, & i \neq j\\
    \end{array} \right. \nonumber \\
    &=& 
    \left\{ \begin{array}{ll}
        \lr{\mb{a}_i^H}\lr{\mb{a}_i} + \tr\{\bs{\Sigma_{\mb{a}_i}}\}, &  i = j \\
        \lr{\mb{a}_i^H}\lr{\mb{a}_j}, & i\neq j,\\
    \end{array} \right.
\end{eqnarray}
which results in:
\begin{align}
\label{eq:eq_num3}
    \lr{\mb{A}^H\mb{A}} = \lr{\mb{A}}^H\lr{\mb{A}} + \mb{D}.
\end{align}

Then, we use~\eqref{eq:eq_num2} and \eqref{eq:eq_num3} to obtain the following expression:
\begin{align}
\label{eq:eq_num4}
    &\tr\big\{\lr{\mb{A}^H\mb{A}}\lr{\mb{x}\mb{x}^H}\big\} = \tr\big\{\lr{\mb{A}^H}\lr{\mb{A}}\lr{\mb{x}}\lr{\mb{x}^H}\big\} + \lr{\mb{x}}^H\mb{D}\lr{\mb{x}} \nonumber \\
    & \quad \quad \quad \quad \quad \quad \quad + \tr\big\{\hat{\bs{\Sigma}}_{\mb{x}}\mb{D}\big\}  +  \tr\big\{\hat{\bs{\Sigma}}_{\mb{x}}\lr{\mb{A}^H}\lr{\mb{A}}\big\}.
\end{align}

Finally, we apply~\eqref{eq:eq_num4} to~\eqref{expand} and remove the duplicated terms to prove~\eqref{f-ABC}.
\end{IEEEproof}
	
\begin{corollary} \label{corol_1}
Given $\mb{x}$ is deterministic, $\blr{\|\mb{y}-\mb{A}\mb{x}\|^2}$ is simplified to:
    \begin{eqnarray}\label{f-A}
        \blr{\|\mb{y}-\mb{A}\mb{x}\|^2} = \|\mb{y}-\lr{\mb{A}}{\mb{x}}\|^2 
        + \sum_{i=1}^n |x_i|^2 \tr\{\hat{\bs{\Sigma}}_{\mb{a}_i}\}.
    \end{eqnarray}
\end{corollary}
\begin{IEEEproof}
    We derive~\eqref{f-A} by setting $\hat{\bs{\Sigma}}_{\mb{x}} = \mb{0}$ and  ${\mb{x}}^H\mb{D}\mb{x}  =   \sum_{i=1}^n |x_i|^2 \tr\{\hat{\bs{\Sigma}}_{\mb{a}_i}\}$ in Lemma~\ref{theorem-2}.
\end{IEEEproof}


\subsubsection{ Updating $x_{i,t}$} We only use this update when $T_p < t \leq T$. In this part, we take the expectation of~\eqref{conditional} with respect to all latent variables except for $x_{i,t}$ to find the variational distribution $q_i(x_{i,t})$ as below:
\begin{align} 
    &q_i(x_{i,t}) \nonumber \\
    &\propto \mr{exp}\Big\{\blr{\ln p(\mb{y}_t|\mb{x}_t,\mb{H}_t,\gamma_t) + \ln p(x_{i,t})}\Big\} \nonumber\\
    &\propto  p(x_{i,t})\, \mr{exp}\big\{\blr{-\gamma_t \big\|\mb{y}_t-\mb{H}_t\mb{x}_t\big\|^2}\big\} \nonumber \\
    &\propto p(x_{i,t})\, \mr{exp}\bigg\{\!-\!\lr{\gamma_t} \bblr{\bigg\|\mb{y}_t-\mb{h}_{i,t}x_{i,t} - \sum_{j\neq i}^K \mb{h}_{j,t}x_{j,t}\bigg\|^2}\bigg\} \nonumber\\
    &\propto p(x_{i,t})\,\mr{exp}\bigg\{\!-\!\lr{\gamma_t} \bigg[ \blr{\|\mb{h}_{i,t}\|^2}|x_{i,t}|^2 \nonumber \\
    & \quad\quad\quad - 2 \,\Re\bigg\{\!\blr{\mb{h}_{i,t}^H}\bigg(\mb{y}_t - \sum_{j\neq i}^K\!\blr{\mb{h}_{j,t}} \lr{x_{j,t}} \bigg)x_{i,t}^*\bigg\}  \bigg]\bigg\} \nonumber \\
    &\propto  p(x_{i,t})\,\mr{exp}\big\{\! -\!\lr{\gamma_t} \blr{\|\mb{h}_{i,t}\|^2} |x_{i,t}-z_{i,t}|^2\big\}, \label{x-t-i}
\end{align}
where 
\begin{align}
\label{eq:z_i,t}
z_{i,t}\triangleq\frac{\lr{\mb{h}_{i,t}^H}}{\blr{\|\mb{h}_{i,t}\|^2}}\Bigg(\mb{y}_t - \sum_{j\neq i}^K \lr{\mb{h}_{j,t}}\lr{x_{j,t}}\Bigg),
\end{align}
acts as a linear estimate of $x_{i,t}$. In~\eqref{eq:z_i,t}, we use Corollary~\ref{corol_1} to compute $\big\langle\|\mb{h}_{i,t}\|^2\big\rangle$ as follows:
\begin{align}
    \big\langle\|\mb{h}_{i,t}\|^2\big\rangle = \|\langle\mb{h}_{i,t}\rangle\|^2 + \tr\{\hat{\bs{\Sigma}}_{i,t}\}.
\end{align}

Since the prior distribution $p(x_{i,t})$ is discrete, the variational distribution $q_i(x_{i,t})$ is also discrete. Therefore, we need to normalize it such that:
\begin{align} 
q_i(a) = \frac{p_a\,\mr{exp}\big\{\! -\!\lr{\gamma_t} \blr{\|\mb{h}_{i,t}\|^2}|a-z_{i,t}|^2 \big\}}
{\sum_{b\in\mc{S}} p_b\,\mr{exp}\big\{\! -\!\lr{\gamma_t} \blr{\|\mb{h}_{i,t}\|^2}|a-z_{i,t}|^2 \big\}},\forall a\in\mc{S}.
\end{align} 

As a result, the variational mean and variance of $x_{i,t}$ are equal to:
\begin{align}
\label{eq:mean_x_i_t}
	\lr{x_{i,t}} &= \sum_{a\in\mc{S}} a q_i(a),\\
\label{eq:var_x_i_t}
 \tau^x_{i,t} &= \sum_{a\in\mc{S}} |a|^2 q_i(a) - |\lr{x_{i,t}}|^2. 
\end{align}


\subsubsection{ Updating $\gamma_t$} In the last part of the CAVI algorithm, we compute $q(\gamma_t$). To achieve this, we take the expectation of~\eqref{conditional} with respect to all latent variables except for $\gamma_t$, to derive the variational distribution $q(\gamma_t)$ as follows:
\begin{align} 
\label{eq:q(gamma_t)}
    q(\gamma_t) &\propto \mr{exp}\Big\{\big\langle\ln p(\mb{y}_t|\mb{x}_t,\mb{H}_t,\gamma_t) + \ln p(\gamma_t) \big\rangle\Big\}  \nonumber \\
    &\propto \mr{exp}\Big\{M\ln \gamma_t - \gamma_t \blr{\|\mb{y}_t-\mb{H}_t\mb{x}_t\|^2} \nonumber \\
    &\quad\quad\quad\quad + (a_0-1)\ln\gamma_t - b_0\gamma_t\Big\}.
\end{align}

Based on~\eqref{eq:q(gamma_t)}, $q(\gamma_t)$ is Gamma distribution with mean
\begin{eqnarray}\label{gamma-t}
    \lr{\gamma_t} = \frac{a_0 + M}{b_0 + \blr{\|\mb{y}_t-\mb{H}_t\mb{x}_t\|^2}},
\end{eqnarray}
where $
\blr{\|\mb{y}_t-\mb{H}_t\mb{x}_t\|^2} = \|\mb{y}_t-\lr{\mb{H}_t}\lr{\mb{x}_t}\|^2 + \sum\limits_{i=1}^K\big[  \tau^x_{i,t}\|\lr{\mb{h}_{i,t}}\|^2 + \lr{|x_{i,t}|^2}\tr\{\bs{\Sigma}_{i,t}\} \big]$ using Lemma \ref{theorem-2}.
	
\begin{figure*}[ht]
\vspace{-5mm}
\centering
  \includegraphics[trim = 0mm 0mm 0mm 0mm, clip, scale=5, width=0.65\linewidth, draft=false]{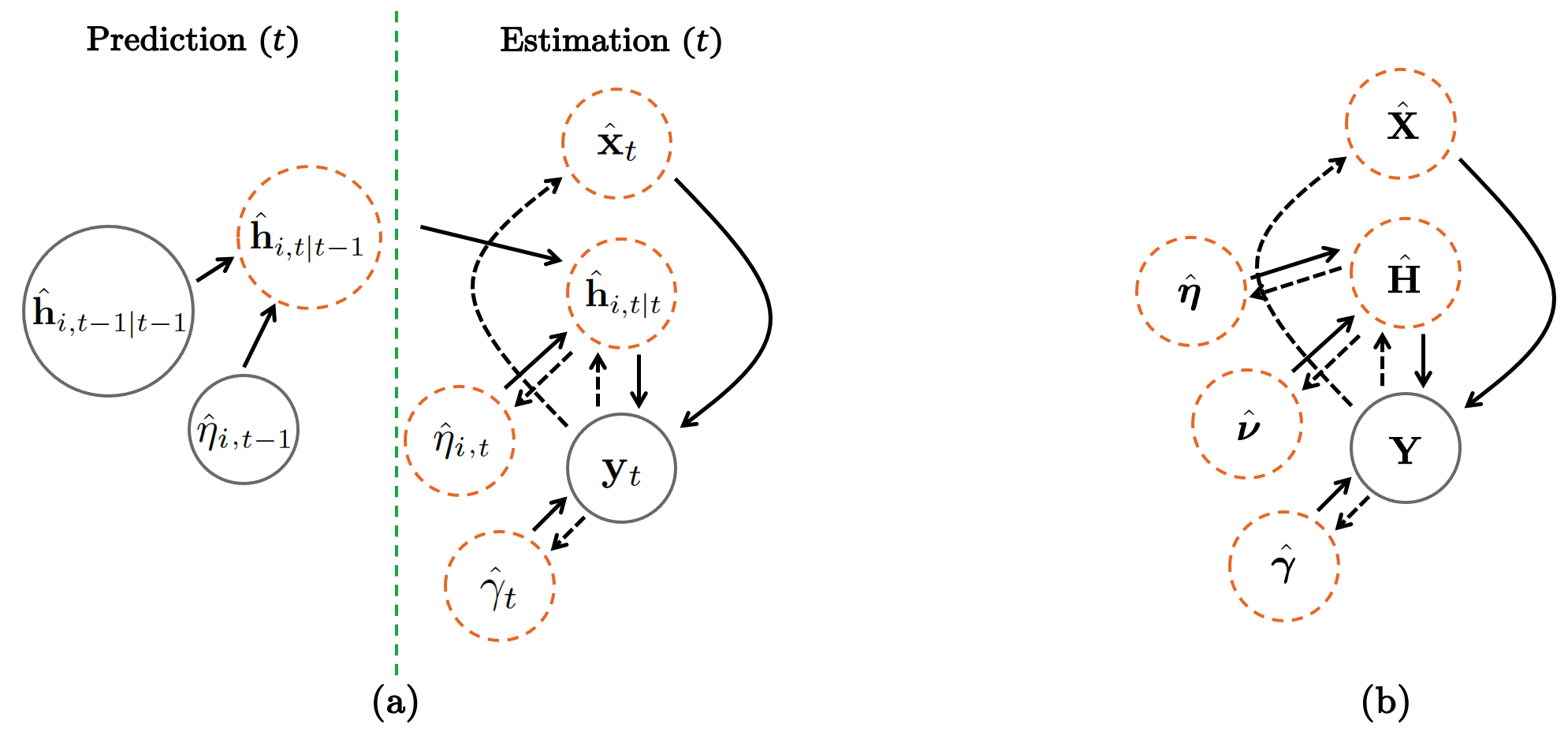}
  \vspace{-2mm}
  \caption{Block diagram of (a) online processing and (b) block processing strategies.}\label{Fig:online_vs_block_diagram}
  \vspace{-4mm}
\end{figure*}

The block diagram of the online processing strategy is shown in Fig. \ref{Fig:online_vs_block_diagram}(a), where we use solid gray circles to denote variables (parameters) that are either known or can be obtained through prior distribution. Additionally, we use dashed orange circles to denote variables that need to be estimated.

Note that the Gaussian variational distributions of $\mb{h}_{i,t}$ and $\eta_i$, along with the Gamma variational distribution of $\gamma_t$, justify the three assumptions made earlier in Table~\ref{table:prior_dist_online}. In Algorithm~\ref{algo-1}, we demonstrate the details of our online processing strategy using the CAVI algorithm. Here, we use $\mb{Y}=[\mb{y}_1, \mb{y}_2, \ldots, \mb{y}_T]$, $\mb{X}=[\mb{x}_1, \mb{x}_2, \ldots, \mb{x}_T]$, and $\mb{H}=[\mb{H}_1, \mb{H}_2, \ldots, \mb{H}_T]$. We define the convergence condition to occur when the number of iterations reaches its maximum value $I_{\mathrm{tr}}$.

As discussed, the online processing strategy focuses only on the information available at time $t$, which leads to a low-complexity low-latency JED approach. However, due to its limited information, it cannot fully eliminate errors in channel estimation and data detection, resulting in error propagation in the JED process. To mitigate JED errors, we propose block processing, which is explained in detail in the next section.

\begin{algorithm}[t]
    \small
    \textbf{Input:} $\mb{Y}$, $\hat{\mb{h}}_{i,0|0}$, $\hat{\bs{\Sigma}}_{i,0|0}$, prior distribution $p_a, \forall a \in \mc{S}$, prior distribution $p(\eta_i)$ (i.e., $\hat{\eta}_{i,0}$ and $\tau^{\eta}_{i,0}$), prior distribution $p(\gamma_t)$ (i.e., $a_0$ and $b_0$), and $I_\mathrm{tr}$\;
    \textbf{Output:} $\mb{X}$, $\mb{H}$, $\bs{\eta}$, and $\gamma_{t}$\; 
    \SetAlgoNoLine
    \For{$t=1,2,\ldots,T$}{
    \If{$T_p < t \leq T$}{
        Initialize $\lr{\mb{x}_{t}} = \mb{0}$;\\
        }
    \For{$i=1,2,\ldots,K$}{
        Use~\eqref{h-prior-mean} and \eqref{h-prior-cov} to predict the prior distribution of $\mb{h}_{i,t}$;\\
        Set $\lr{\mb{h}_{i,t}} =\hat{\mb{h}}_{i,t|t-1}$, $\bs{\Sigma}_{i,t} = \hat{\bs{\Sigma}}_{i,t|t-1}$;\\
        Compute $\lr{\eta_i}$ and $\lr{\gamma_{t}}$ based on the prior distributions of $\eta_i$ and $\gamma_t$\;	
        }
        \Repeat{convergence}{
            \For{$i=1,2,\ldots,K$}{
                Find the distribution $q(\mb{h}_{i,t})$ using~\eqref{q-h-i}\;
                Obtain $\bs{\Sigma}_{i,t}$ and $\lr{\mb{h}_{i,t}}$ using~\eqref{Sigma-h} and~\eqref{mean-h}\;}
            \For{$i=1,2,\ldots,K$}{
                Attain the distribution $q(\eta_i)$ as in~\eqref{eta-i}\;
                Determine $\tau^{\eta}_{i,t}$ and $\lr{\eta_i}$ using~\eqref{tau_i_var} and~\eqref{tau_i_mean}\;}
            \If{$T_p < t \leq T$}{
                \For{$i=1,2,\ldots,K$}{
                    Get the distribution $q_i(x_{i,t})$ as in \eqref{x-t-i}\;
                    Calculate $\lr{x_{i,t}}$ and $\tau^x_{i,t}$ based on~\eqref{eq:mean_x_i_t} and~\eqref{eq:var_x_i_t}\;}
            }
            Compute $\lr{\gamma_{t}}$ as in \eqref{gamma-t};}
    }
        Set $\hat{\mb{h}}_{i,t|t} = \lr{\mb{h}_{i,t}}$, $\hat{\bs{\Sigma}}_{i,t|t} = \bs{\Sigma}_{i,t}$, $\hat{\eta}_{i,t} = \lr{\eta_i}$, and $\gamma_t = \lr{\gamma_t}$\;
    
Calculate $\hat{x}_{i,t} = \argmax_{a\in \mc{S}} q_i(a)$.
    \caption{The Online Processing Strategy}
    \label{algo-1}
\end{algorithm}

%% file: Sections/VB_for_JED_Block_Processing.tex
In real-world massive MIMO systems, the BS typically estimates uplink channels and uses them for downlink beamforming to enhance data rates and quality of service (QoS), making accurate channel estimation essential. In this section, we propose a block processing strategy to reduce channel estimation errors. The concept of block processing originates from estimation theory, where the estimation error decreases as more observed signals become available. We first formulate the problem, where the communication block spans the entire communication time, stacking all received signals from time $1$ to $T$ and processing them together. Next, we introduce a scheme for acquiring Bayesian optimal estimates of $\mb{H}_t$ and $\mb{x}_t$, and then apply the CAVI algorithm to sequentially update variables to obtain a local optimal solution.
\subsection{Problem Formulation}
Similar to the online strategy in Section~\ref{Sec:VB_for_JED}, we assume $\eta_i$ and $\gamma_t$ are i.i.d. unknown variables. Then, we compute joint distribution $p(\mb{Y},\mb{X},\mb{H}_1, \ldots,\mb{H}_{T},\bs{\gamma},\bs{\eta};\bar{\mb{R}})$ as below:
\begin{align}
\label{eq:joint_prob_block}
    &p(\mb{Y},\mb{X},\mb{H}_1,\ldots,\mb{H}_{T},\bs{\gamma},\bs{\eta};\bar{\mb{R}}) = \left[ \prod_{t=1}^{T} p(\mb{H}_{t}|\mb{H}_{t-1}, \bs{\eta};\mb{R}_i) \right] \nonumber \\
    & \quad \quad~~\times \left[\prod_{t=1}^{T}p(\mb{y}_t|\mb{x}_t,\mb{H}_t,\gamma_t)p(\mb{x}_t) p(\gamma_t)\right] p(\bs{\eta})p(\mb{H}_{0}),
\end{align}
where 
$\bs{\gamma}=[\gamma_1, \gamma_2,\ldots,\gamma_T]$ and $p(\mb{H}_0) = \prod_{i=1}^{K}p(\mb{h}_{i,0})$ is the prior channel distribution. In this case, to compute the conditional probability  $p(\mb{h}_{i,t}|\mb{h}_{i,t-1},\eta_i;\mb{R}_i)$, we use~\eqref{eq_cond_ht_ht_1} and rewrite it as:
\begin{align}
\label{eq:cond_prob_nu}
    p(\mb{h}_{i,t}|\mb{h}_{i,t-1},\eta_i;\mb{R}_i)
    &=\mc{CN}\left(\mb{h}_{i,t};\eta_i\mb{h}_{i,t-1},  \nu_i^{-1}\mb{R}_i\right),
    \end{align}
where $\nu_i \stackrel{\triangle}{=} (1-\eta_i^2)^{-1}$. In the following subsection, we apply the VB framework to estimate $\nu_i$.

\subsection{Estimation phase}
In this part, we focus on the entire communication block for JED. Unlike online processing, we simplify the process by bypassing the prediction phase, and then we solely rely on the estimation phase to derive the Bayesian optimal estimates for $\mathbf{H}_t$ and $\mathbf{x}_t$. We achieve this through the posterior distribution $p\left(\mb{X},\mb{H}_1,\ldots, \mb{H}_T, \bs{\gamma}, \bs{\eta}|\mb{Y}\right)$. Although~\eqref{eq:cond_prob_nu} shows that $\nu_i$ and $\eta_i$ are dependent, in the estimation process, we assume these two variables are independent and hope to estimate $\nu_i$ such that $\lr{\nu_i} = \left(1- \lr{\eta_i}^2\right)^{-1}$. Our simulation results in Section \ref{Sec:numerical} will validate this assumption. Therefore, we replace the posterior distribution $p\left(\mb{X},\mb{H}_1,\ldots, \mb{H}_T, \bs{\gamma}, \bs{\eta}|\mb{Y}\right)$ by $p\left(\mb{X},\mb{H}_1,\ldots, \mb{H}_T, \bs{\gamma}, \bs{\eta},\bs{\nu}|\mb{Y}\right)$, where $\bs{\nu} = [\nu_1,\nu_2,\ldots, \nu_K]$. Then, we use the mean-field variational distribution $q\left(\mb{X},\mb{H}_1,\ldots, \mb{H}_T, \bs{\gamma}, \bs{\eta},\bs{\nu}\right)$ to approximate $p\left(\mb{X},\mb{H}_1,\ldots, \mb{H}_T, \bs{\gamma}, \bs{\eta},\bs{\nu}|\mb{Y}\right)$, which is given by:
\begin{align} 
    \label{eq:mean_field_block}
&p\left(\mb{X},\mb{H}_1,\ldots, \mb{H}_T, \bs{\gamma}, \bs{\eta},\bs{\nu}|\mb{Y}\right) \approx q\left(\mb{X},\mb{H}_1,\ldots, \mb{H}_T, \bs{\gamma}, \bs{\eta},\bs{\nu}\right) \nonumber \\ 
& \quad \quad \quad \quad \quad ~~ = \prod_{i=1}^{K} \prod_{t=1}^{T}q_i(x_{i,t}) 
    q(\mb{h}_{i,t}) q(\eta_i) q(\nu_i)q(\gamma_t).
\end{align}

Further, referring to~\eqref{eq:q_start_prop}, the optimal solution for the variational densities in \eqref{eq:mean_field_block} requires the joint distribution $p(\mb{Y},\mb{X},\mb{H}_1,\ldots,\mb{H}_{T},\bs{\gamma},\bs{\eta}, \bs{\nu};\bar{\mb{R}})$. To obtain this, one can factorize $p(\mb{Y},\mb{X},\mb{H}_1,\ldots,\mb{H}_{T},\bs{\gamma},\bs{\eta}, \bs{\nu};\bar{\mb{R}})$ as follows:
\begin{align}
    \label{conditional_block}
        &p(\mb{Y},\mb{X},\mb{H}_1,\ldots,\mb{H}_{T},\bs{\gamma},\bs{\eta}, \bs{\nu};\bar{\mb{R}}) = \left[ \prod_{i=1}^{K} p(\eta_i) p(\mb{h}_{i,0})\right] \nonumber \\
        &\quad \quad \quad \times \left[ \prod_{i=1}^{K} p(\nu_i)\right]\left[ \prod_{i=1}^{K} \prod_{t=1}^{T} p(\mb{h}_{i,t}|\mb{h}_{i,t-1}, \eta_i, \nu_i;\mb{R}_i) \right]   \nonumber \\
        &\quad \quad \quad \times  \left[\prod_{t=1}^{T}p(\mb{y}_t|\mb{x}_t,\mb{H}_t,\gamma_t)\left[\prod_{i=1}^{K}p(x_{i,t})\right] p(\gamma_t)\right].
\end{align}

In this strategy, we need the prior probabilities $p(\eta_i)$, $p(\gamma_t)$, $p(\mb{h}_{i,t}|\mb{h}_{i,t-1}, \eta_i, \nu_i; \mb{R}_i)$, $p(x_{i,t})$, and $p(\nu_i)$ to attain the variational distributions in~\eqref{eq:mean_field_block}, which are outlined in Table~\ref{table:prior_dist_block}. We show the block diagram of the block processing strategy in Fig.~\ref{Fig:online_vs_block_diagram}(b).

\begin{table} [ht]
\caption{The prior distribution assumptions for the unknown random variables involved in $q\left(\mb{X},\mb{H}_1,\ldots, \mb{H}_T, \bs{\gamma}, \bs{\eta},\bs{\nu}\right) $
}
\centering
{
\begin{tabular}{|c|c|}
\hline
Probability & Prior Distribution Assumption\\
\hline \hline
$p(\mb{h}_{i,t}|\mb{h}_{i,t-1}, \eta_i, \nu_i; \mb{R}_i)$ & $\mc{CN}\left(\mb{h}_{i,t};\eta_i\mb{h}_{i,t-1},  \nu_i^{-1}\mb{R}_i\right)$ \\

$p(\gamma_t)$ & $\Gamma(a_{0,b},b_{0,b})$ \\

$p(\eta_i)$ & $\mc{N}\big(\eta_i; \hat{\eta}_{i,b}, \tau^{\eta}_{i,b}\big)$ \\

$p(\nu_i)$ & $\Gamma(\bar{a}_{i,b},\bar{b}_{i,b})$\\

$p(x_{i,t}) (\text{pilot transmission})$ & $\delta(x_{i,t} - \bar{x}_{i,t})$\\

$p(x_{i,t}) (\text{data transmission})$ & $\sum_{a\in \mc{S}} p_a\delta(x_{i,t} - a)$\\

\hline
\end{tabular}
}
\label{table:prior_dist_block}
\end{table}





\subsection{The CAVI algorithm}
Here, akin to the approach used in online processing, we apply the iterative CAVI algorithm to determine the local optimal solutions for $q^{\star}(x_{i,t})$, $q^{\star}(\mb{h}_{i,t})$, $q^{\star}(\eta_i)$, $q^{\star}(\gamma_t)$, and $q^{\star}(\nu_i)$. It is important to note that the updating process for $x_{i,t}$ and $\gamma_t$ closely resembles the procedure employed in the online processing strategy. Thus, we only provide the specifics of updating $\mb{h}_{i,t}$, $\eta_i$, and $\nu_i$ in this subsection.

\subsubsection{ Updating $\mb{h}_{i,t}$} 

In Section~\ref{Sec:VB_for_JED}, 
we compute $q(\mb{h}_{i,t})$ based on the predictive distribution and the conditional probability $p(\mb{y}_t|\mb{H}_t, \mb{x}_t, \gamma_t)$ at time $t$. However, as shown in~\eqref{conditional_block}, the block processing strategy requires three conditional probabilities that have $\mb{h}_{i,t}$ as a factor, i.e., $p(\mb{y}_t|\mb{H}_t, \mb{x}_t, \gamma_t)$, $p(\mb{h}_{i,t}|\mb{h}_{i,t-1}, \eta_i, \nu_i; \mb{R}_i)$, and $p(\mb{h}_{i,t+1}|\mb{h}_{i,t}, \eta_i, \nu_i; \mb{R}_i)$ to derive
$q(\mb{h}_{i,t})$. By taking the expectation of the conditional probability in~\eqref{conditional_block} with respect to all latent variables except for $\mb{h}_{i,t}$, the variational distribution $q(\mb{h}_{i,t})$ is given in \eqref{q-h-i_block} (at the top of the next page), which  is Gaussian with covariance $\bs{\Sigma}_{i,t}$ and mean $\lr{\mb{h}_{i,t}}$, as outlined below:
\begin{align}
\bs{\Sigma}_{i,t}  &=\Big[\lr{\gamma_t}\lr{|x_{i,t}|^2}\mb{I}_M+ \left(1 +  \lr{\eta_i^2}\right) \lr{\nu_i} \mb{R}^{-1}_{i}\Big]^{-1},  \label{Sigma-h_block} \\
\lr{\mb{h}_{i,t}} &=\bs{\Sigma}_{i,t}\Bigg[\lr{\gamma_t}\bigg(\mb{y}_t- \sum_{j\neq i}^K \lr{\mb{h}_{j,t}}\lr{x_{j,t}}\bigg)\lr{x_{i,t}^*}~+ \nonumber \\
& \quad \quad \quad \quad \lr{\eta_i} \lr{\nu_i} \mb{R}^{-1}_{i}\left(\mb{h}_{i,t+1} + \mb{h}_{i,t-1}\right)\Bigg].  \label{mean-h_block}
\end{align}

	\begin{figure*}
        \begin{align}
        \label{q-h-i_block}
            q(\mb{h}_{i,t}) &\propto \mr{exp}\bigg\{\big\langle\ln p(\mb{y}_t|\mb{x}_t,\mb{H}_t,\gamma_t) + \ln p(\mb{h}_{i,t}|\mb{h}_{i,t-1},\eta_i, \nu_i;\mb{R}_i) + \ln p(\mb{h}_{i,t+1}|\mb{h}_{i,t},\eta_i, \nu_i;\mb{R}_i) \big\rangle\bigg\}  \nonumber \\
            &\propto \mr{exp}\bigg\{- \blr{\gamma_t \big\|\mb{y}_t-\mb{H}_t\mb{x}_t\big\|^2} -  \bblr{ (\mb{h}_{i,t}-\eta_i\mb{h}_{i,t-1})^H \nu_i \mb{R}^{-1}_{i}(\mb{h}_{i,t}-\eta_i\mb{h}_{i,t-1})} \nonumber \\
            & \quad \quad - \bblr{ (\mb{h}_{i,t+1}-\eta_i\mb{h}_{i,t})^H \nu_i \mb{R}^{-1}_{i}(\mb{h}_{i,t+1}-\eta_i\mb{h}_{i,t})}\bigg\} \nonumber \\
            &\propto \mr{exp}\bigg\{- \bblr{\gamma_t \bigg\|\mb{y}_t - \mb{h}_{i,t}x_{i,t} - \sum_{j\neq i}^K \mb{h}_{j,t}x_{j,t}\bigg\|^2} -  \bblr{ (\mb{h}_{i,t}-\eta_i\mb{h}_{i,t-1})^H \nu_i \mb{R}^{-1}_{i}(\mb{h}_{i,t}-\eta_i\mb{h}_{i,t-1})} \nonumber \\
            & \quad \quad - \bblr{ (\mb{h}_{i,t+1}-\eta_i\mb{h}_{i,t})^H \nu_i \mb{R}^{-1}_{i}(\mb{h}_{i,t+1}-\eta_i\mb{h}_{i,t})}\bigg\} \nonumber \\
            &\propto \mr{exp}\Bigg\{-\mb{h}_{i,t}^H\Big[\lr{\gamma_t}\lr{|x_{i,t}|^2}\mb{I}_M+ \left(1 +  \lr{\eta_i^2}\right)  \lr{\nu_i} \mb{R}^{-1}_{i}\Big]\mb{h}_{i,t}  \nonumber \\
			&\quad\quad + 2\,\Re\bigg\{\lr{\gamma_t}\mb{h}_{i,t}^H\bigg(\mb{y}_t- \sum_{j\neq i}^K \lr{\mb{h}_{j,t}}\lr{x_{j,t}}\bigg)\lr{x_{i,t}^*} + \lr{\eta_i} \lr{\nu_i} \mb{h}_{i,t}^H\mb{R}^{-1}_{i}\left(\mb{h}_{i,t+1} + \mb{h}_{i,t-1}\right)\bigg\}\Bigg\}.
        \end{align}
      \hrulefill
      \vspace{-3mm}
	\end{figure*}
\subsubsection{ Updating $\eta_i$} Taking the expectation of the conditional probability in~\eqref{conditional_block} with respect to all latent variables except for $\eta_i$, we obtain the variational distribution $q(\eta_i)$ as presented in~\eqref{eta-i_block} (on the next page). Specifically, to derive $q(\eta_i)$, we require the conditional channel probabilities throughout the entire block and the prior probability of $p\big(\eta_i;\hat{\eta}_{i,b},\tau^{\eta}_{i,b}\big)$. Notice $\hat{\eta}_{i,b} = \hat{\eta}_{i,0}$ and $\tau^{\eta}_{i,b} = \tau^{\eta}_{i,0}$ in~\eqref{eta-i_block}.
\begin{figure*}
  \begin{eqnarray} 
  \label{eta-i_block}
			q(\eta_i) &\propto& \mr{exp}\Big\{\blr{ \sum_{t=1}^{T} \ln p\big(\mb{h}_{i,t}|\mb{h}_{i,t-1},\eta_i, \nu_i;\mb{R}_i\big) + \ln p\big(\eta_i;\hat{\eta}_{i,t-1},\tau^{\eta}_{i,t-1}\big)}\Big\}   \\
			&\propto& \mr{exp}\Big\{ - \Blr{\sum_{t=1}^{T}\big(\mb{h}_{i,t}-\eta_i\mb{h}_{i,t-1}\big)^H \nu_i \mb{R}_{i}^{-1}\big(\mb{h}_{i,t}-\eta_i\mb{h}_{i,t-1}\big)}  - \|\eta_i-\hat{\eta}_{i,t-1}\|^2 / \tau^{\eta}_{i,t-1}\Big\} \nonumber \\
			&\propto& \mr{exp}\Bigg\{ - \eta_i^2\Big(\lr{\nu_i}\sum_{t=1}^{T}\lr{\mb{h}_{i,t-1}}^H\mb{R}_{i}^{-1}\lr{\mb{h}_{i,t-1}} + 1/\tau^{\eta}_{i,t-1}\Big) +  2\eta_i\Big(\Re\big\{\lr{\nu_i}\sum_{t=1}^{T}\lr{\mb{h}_{i,t-1}}^H\mb{R}_{i}^{-1}\lr{\mb{h}_{i,t}} \big\} + \frac{\hat{\eta}_{i,t-1}}{\tau^{\eta}_{i,t-1}}\Big) \Bigg\} \nonumber,
		\end{eqnarray}
  \hrulefill
  \vspace{-3mm}
  \end{figure*}
Here, the variational distribution $q(\eta_i)$ is Gaussian with the following variance and mean:
\begin{align}
\label{tau_i_var_block}
	\tau^{\eta}_{i,t} &= \Big(\lr{\nu_i}\sum_{t=1}^{T}\lr{\mb{h}_{i,t-1}}^H\mb{R}_{i}^{-1}\lr{\mb{h}_{i,t-1}} + 1/\tau^{\eta}_{i,t-1}\Big)^{-1},\\
 \label{tau_i_mean_block}
		\lr{\eta_{i,t}} &= \tau^{\eta}_{i,t}  \Big(\Re\big\{\lr{\nu_i}\sum_{t=1}^{T}\lr{\mb{h}_{i,t-1}}^H\mb{R}_{i}^{-1}\lr{\mb{h}_{i,t}} \big\} + \frac{\hat{\eta}_{i,t-1}}{\tau^{\eta}_{i,t-1}}\Big).
	\end{align}
\subsubsection{ Updating $\nu_i$} 
In this case, we assume $\nu_i$ is independent of $\eta_i$ and subsequently employ the VB framework to estimate the variational distribution $q(\nu_i)$. To derive $q(\nu_i)$, the following lemma is essential: 

\begin{lemma}
\label{lemma-2-q(nu_i)}
\cite{nguyen2022variationaladc} Let $\mb{A} \in \mathbb{C}^{m\times n}$, $\mb{x}\in \mathbb{C}^{n\times 1}$, and $\mb{y}\in \mathbb{C}^{m\times 1}$ be three independent random matrices (vectors) with respect to a variational density $q_{\mb{A},\mb{x},\mb{y}}(\mb{A},\mb{x},\mb{y}) = q_{\mb{A}}(\mb{A})q_{\mb{x}}(\mb{x})q_{\mb{y}}(\mb{y})$. Suppose $\mb{A}$ is column-wise independent and let $\lr{\mb{a}_i}$ and $\bs{\Sigma}_{\mb{a}_i}$ denote the variational mean and covariance matrix of the $i^{th}$ column of $\mb{A}$. Let $\lr{\mb{x}}$ and $\bs{\Sigma}_{\mb{x}}$ represent the variational mean and covariance matrix of $\mb{x}$, respectively. Further, assume $\lr{\mb{y}}$ and $\bs{\Sigma}_{\mb{y}}$ are the variational mean and covariance matrix of $\mb{y}$, respectively. Then, for an arbitrary Hermitian matrix $\mb{R}$, we define $\blr{\left(\mb{y}-\mb{A}\mb{x}\right)^H \mb{R} \left(\mb{y}-\mb{A}\mb{x}\right)}$ as the expectation of $\left(\mb{y}-\mb{A}\mb{x}\right)^H \mb{R} \left(\mb{y}-\mb{A}\mb{x}\right)$ with respect to $q_{\mb{A},\mb{x},\mb{y}}(\mb{A},\mb{x},\mb{y})$. Here, we have:
\begin{align}\label{f-ABCD}
    &\blr{\left(\mb{y}-\mb{A}\mb{x}\right)^H \mb{R} \left(\mb{y}-\mb{A}\mb{x}\right)} = \lr{\mb{x}}^H\mb{D}\lr{\mb{x}} + \tr\big\{\bs{\Sigma}_{\mb{x}}\mb{D}\big\} \nonumber \\
    & \quad \quad \quad \quad \quad + \left(\blr{\mb{y}}-\lr{\mb{A}} \lr{\mb{x}}\right)^H \mb{R} \left(\lr{\mb{y}}-\lr{\mb{A}} \lr{\mb{x}}\right) 
      \nonumber \\ 
      &\quad \quad \quad \quad \quad +\tr\big\{\mb{R}\bs{\Sigma}_{\mb{y}}\big\} + \tr\big\{\bs{\Sigma}_{\mb{x}}\lr{\mb{A}^H} \mb{R} \lr{\mb{A}} \big\},
\end{align}
where $\mb{D} = \mr{diag}\big(\tr\{\mb{R}\bs{\Sigma_{\mb{a}_1}}\},\ldots, \tr\{\mb{R}\bs{\Sigma_{\mb{a}_n}}\}\big)$. 
\end{lemma}
\begin{IEEEproof}
We omit the proof of this lemma since it is similar to the proof of Lemma~\ref{theorem-2}. 
\end{IEEEproof}

By taking the expectation of the probability in~\eqref{conditional_block} with respect to all latent variables except for $\nu_i$, we have:
\begin{align} 
  \label{nu-i_block}
    q(\nu_i) &\propto \mr{exp}\Big\{\blr{ \sum_{t=1}^{T} \ln p\big(\mb{h}_{i,t}|\mb{h}_{i,t-1},\eta_i, \nu_i;\mb{R}_i\big) \\
    & \quad \quad \quad \quad \quad + \ln p\big(\nu_i;\bar{a}_{i,t-1},\bar{b}_{i,t-1}\big)}\Big\}  \nonumber \\
    &\propto \mr{exp}\Big\{T M \ln \nu_i + \left(\bar{a}_{i,t-1} - 1\right)\ln \nu_i - \bar{b}_{i,t-1} \nu_i \nonumber \\
    &- \nu_i\Blr{\sum_{t=1}^{T}\big(\mb{h}_{i,t}-\eta_i\mb{h}_{i,t-1}\big)^H  \mb{R}_{i}^{-1}\big(\mb{h}_{i,t}-\eta_i\mb{h}_{i,t-1}\big)}\Big\}.\nonumber
\end{align}

Thus, the variational distribution $q(\nu_i)$ is $\Gamma(\bar{a}_{i,t}, \bar{b}_{i,t})$, where:
\begin{align}
    \label{estimation-a-i_block}
    \bar{a}_{i,t} = \bar{a}_{i,t-1} + TM,
\end{align}
and $\bar{b}_{i,t}$ is given in~\eqref{estimation-b-i_block} (on the next page). Here, $\lr{\nu_i}$ is equal to $\bar{a}_{i,t}/\bar{b}_{i,t}$.
\begin{figure*}
\begin{align}
		\bar{b}_{i,t} &= \bar{b}_{i,t-1} + \Blr{\sum_{t=1}^{T}\big(\mb{h}_{i,t}-\eta_i\mb{h}_{i,t-1}\big)^H  \mb{R}_{i}^{-1}\big(\mb{h}_{i,t}-\eta_i\mb{h}_{i,t-1}\big)} \nonumber \\
		&= \bar{b}_{i,t-1} + \sum_{t=1}^{T}\bigg(\lr{\mb{h}_{i,t}} - \lr{\eta_i}\lr{\mb{h}_{i,t-1}}\bigg)^H\mb{R}_{i}^{-1}\bigg(\lr{\mb{h}_{i,t}}-\lr{\eta_i}\lr{\mb{h}_{i,t-1}}\bigg) + \sum_{t=1}^{T}\tr\big\{\mb{R}_{i}^{-1}\bs{\Sigma}_{i,t}\big\} \nonumber \\
		& + \sum_{t=1}^{T}\tau^{\eta}_{i,t} \lr{\mb{h}_{i,t-1}}^H\tr{\{\mb{R}_{i}^{-1}\}}\lr{\mb{h}_{i,t-1}} + \sum_{t=1}^{T}\tr\big\{\tau^{\eta}_{i,t}\bs{\Sigma}_{i,t-1}\tr{\{\mb{R}_{i}^{-1}\}}\big\}  + \sum_{t=1}^{T}\tr\big\{\bs{\Sigma}_{i,t-1}\lr{\eta_{i,t}}\mb{R}_{i}^{-1}\lr{\eta_{i,t}}\big\} \big\},
  \label{estimation-b-i_block} 
\end{align}
\hrulefill
\vspace{-3mm}
\end{figure*}
\begin{algorithm}[t]
    \small
    \textbf{Input:} $\mb{Y}$, $\hat{\mb{h}}_{i,0|0}$, $\hat{\bs{\Sigma}}_{i,0|0}$, prior distribution $p_a, \forall a \in \mc{S}$, prior distribution $p(\eta_i)$ (i.e., $\hat{\eta}_{i,0}$ and $\tau^{\eta}_{i,0}$), prior distribution $p(\nu_i)$ (i.e., $\bar{a}_{i,0}$ and $\bar{b}_{i,0}$), prior distribution $p(\gamma_t)$ (i.e., $a_0$ and $b_0$), and $I_\mathrm{tr}$\;
    \textbf{Output:} $\mb{X}$, $\mb{H}$, $\bs{\eta}$, $\bs{\nu}$, and $\gamma_{t}$\; 
    \SetAlgoNoLine
    \For{$t=1,2,\ldots,T$}{
    \If{$T_p < t \leq T$}{
        Initialize $\lr{\mb{x}_{t}} = \mb{0}$;\\
        }
    \For{$i=1,2,\ldots,K$}{
    Compute $\lr{\eta_i}$, $\nu_i$, and $\lr{\gamma_{t}}$ based on the prior distributions of $\eta_i$, $\nu_i$, and $\gamma_t$\;	
    }
    }
    \Repeat{convergence}{
    \For{$t=1,2,\ldots,T$}{
    \For{$i=1,2,\ldots,K$}{
    
    \If {$t=1$}{
    Set $\mb{h}_{i,t-1} =\hat{\mb{h}}_{i,0|0}$;\\
    Follow Algorithm~\ref{algo-1} to compute $\mb{h}_{i,t+1}$;\\
    }
    \ElseIf {$t=T$}{
    Set $\mb{h}_{i,t+1} =\lr{\eta_i}\hat{\mb{h}}_{i,0|0}$;\\
    }
    \Else{
    Follow Algorithm~\ref{algo-1} to compute $\mb{h}_{i,t+1}$;\\
    }

    Find the distribution $q(\mb{h}_{i,t})$ using~\eqref{q-h-i_block}\;
    Obtain $\bs{\Sigma}_{i,t}$ and $\lr{\mb{h}_{i,t}}$ using~\eqref{Sigma-h_block} and~\eqref{mean-h_block}\;
                
    }   
    }    
        
            \For{$t=1,2,\ldots,T$}{
            \For{$i=1,2,\ldots,K$}{
                Attain the distribution $q(\eta_i)$ as in~\eqref{eta-i_block}\;
                Determine $\tau^{\eta}_{i,t}$ and $\lr{\eta_i}$ using~\eqref{tau_i_var_block} and~\eqref{tau_i_mean_block}\;}
                }

            \For{$t=1,2,\ldots,T$}{
            \For{$i=1,2,\ldots,K$}{
                Find the distribution $q(\nu_i)$ as in~\eqref{nu-i_block}\;
                Compute $\bar{a}_{i,t}$ and $\bar{b}_{i,t}$ using~\eqref{estimation-a-i_block} and~\eqref{estimation-b-i_block}\;}
                }
                \For{$t=1,2,\ldots,T$}{
            \If{$T_p < t \leq T$}{
                \For{$i=1,2,\ldots,K$}{
                    Get the distribution $q_i(x_{i,t})$ as in \eqref{x-t-i}\;
                    Calculate $\lr{x_{i,t}}$ and $\tau^x_{i,t}$ based on~\eqref{eq:mean_x_i_t} and~\eqref{eq:var_x_i_t}\;}
                    }
            Compute $\lr{\gamma_{t}}$ as in \eqref{gamma-t}\;
            }
            }
        Set $\mb{h}_{i,t} = \lr{\mb{h}_{i,t}}$, $\eta_i = \lr{\eta_i}$, $\nu_i = \lr{\nu_i}$, and $\gamma_t = \lr{\gamma_t}$\;
Get $\hat{x}_{i,t} = \argmax_{a\in \mc{S}} q_i(a)$.
    \caption{The Block Processing Strategy}
    \label{algo-block}
\end{algorithm}
In Algorithm~\ref{algo-block}, we explain the details of how our block processing strategy utilizes the CAVI algorithm to optimize $q(\mb{h}_{i,t})$, $q_i(x_{i,t})$, $q(\eta_i)$, $q(\nu_i)$, and $q(\gamma_t)$.        

%% file: Sections/Numerical_Results.tex
In this section, we evaluate the performance of our proposed VB-based approaches in time-varying channels. We consider a JED problem in a massive MIMO network supporting $K=4$ high-mobility users. Initially, we focus on the online processing strategy and compare its performance with three baselines (i.e., the LMMSE, KF, and EP methods) in terms of SER under two cases: (i) when $\eta_i$ is fixed and (ii) when it is a random variable. Furthermore, we compare the computational complexity and channel NMSE of our proposed online strategy with the benchmarks, where NMSE is given by:
\begin{align}
    \mr{NMSE~(dB)} = 10 \log_{10} \left(\frac{\|\mb{H}-\hat{\mb{H}}\|_{\mr{F}}^2}{\|\mb{H}\|_{\mr{F}}^2}\right).
\end{align}

Subsequently, we use an interleaved structure to enhance the SER performance of the VB-based online strategy. Finally, we assess the performance of both our online and block processing strategies in terms of SER and channel NMSE. 

Throughout this section, we use $I_{\mathrm{tr}} = 50$, $T_p = 8$, and $T_d = 128$ to represent the maximum number of iterations for the CAVI algorithm and the KF and EP methods, the pilot transmission time, and the data transmission time, respectively. 

 We normalize the covariance matrix $\mb{R}_i~\forall i$ such that its diagonal elements are all set to $1/M$, resulting in $\mathbb{E}[|\mb{h}_{i}|^2] = 1$. Subsequently, the noise variance $N_0$ is determined using the signal-to-noise ratio (SNR), given by:
\begin{align}
\mr{SNR} = \frac{\mathbb{E}[\|\mb{H}\mb{x}\|^2]}{\mathbb{E}[\|\mb{n}\|^2]} = \frac{\sum_{i=1}^K  \mr{Tr}(\mb{R}_{i})}{MN_0} = \frac{K}{MN_0}.
\end{align}

To model the correlated channels, we consider the exponential spatial correlation model \cite{loyka2001channel} for each column of $\mathbf{H}_t$, in which the covariance matrix $\mathbf{R}_i$ is set to:
\begin{align}
[\mathbf{R}_i]_{k\ell} = 
\begin{cases} 
\frac{1}{M} \alpha^{k-\ell}, & \text{if } k \geq \ell, \\
\frac{1}{M} \left( \alpha^{\ell-k} \right)^*, & \text{if } k < \ell,
\end{cases}
\end{align}
where $1 \leq k, \ell \leq M$, and $\alpha$ is the (complex) correlation coefficient between neighboring receive antennas.

Further, we use the LMMSE channel estimation method to find $\hat{\mb{h}}_{i,0|0}$ and $\hat{\bs{\Sigma}}_{i,0|0}$. We also initialize $p_a = 1/|\mc{S}|$, $\hat{\eta}_{i,0} = 0.95, \tau^{\eta}_{i,0}= 10^{-3}, \bar{a}_{i,0} = 10^{-4}, \bar{b}_{i,0} = 10^{-4}, a_0 = 10^{-4}$, and $b_0 = 10^{-4}$, where $|\mc{S}|$ denotes the cardinality of $\mc{S}$. Ultimately, we evaluate the simulation results across $1000$ trials.

\subsection{Performance of Online Processing VB}
In this part, we evaluate the effectiveness of the proposed online strategy within the VB framework. To accomplish this, we use the LMMSE, KF, and EP methods~\cite{naraghi2021semiICC} as benchmarks and compare them with our approach. Also, to gain deeper insights into the performance of the online VB method, we employ another benchmark where the VB method knows $\eta_i$.

We first focus on the SER metric and examine two cases: one where $\eta_i$ remains constant and another where it varies as a random variable with minor fluctuations.


\begin{figure}[ht]
\vspace{-3mm}
\centering
  \includegraphics[trim = 0mm 0mm 0mm 0mm, clip, scale=9, width=0.99\linewidth, draft=false]{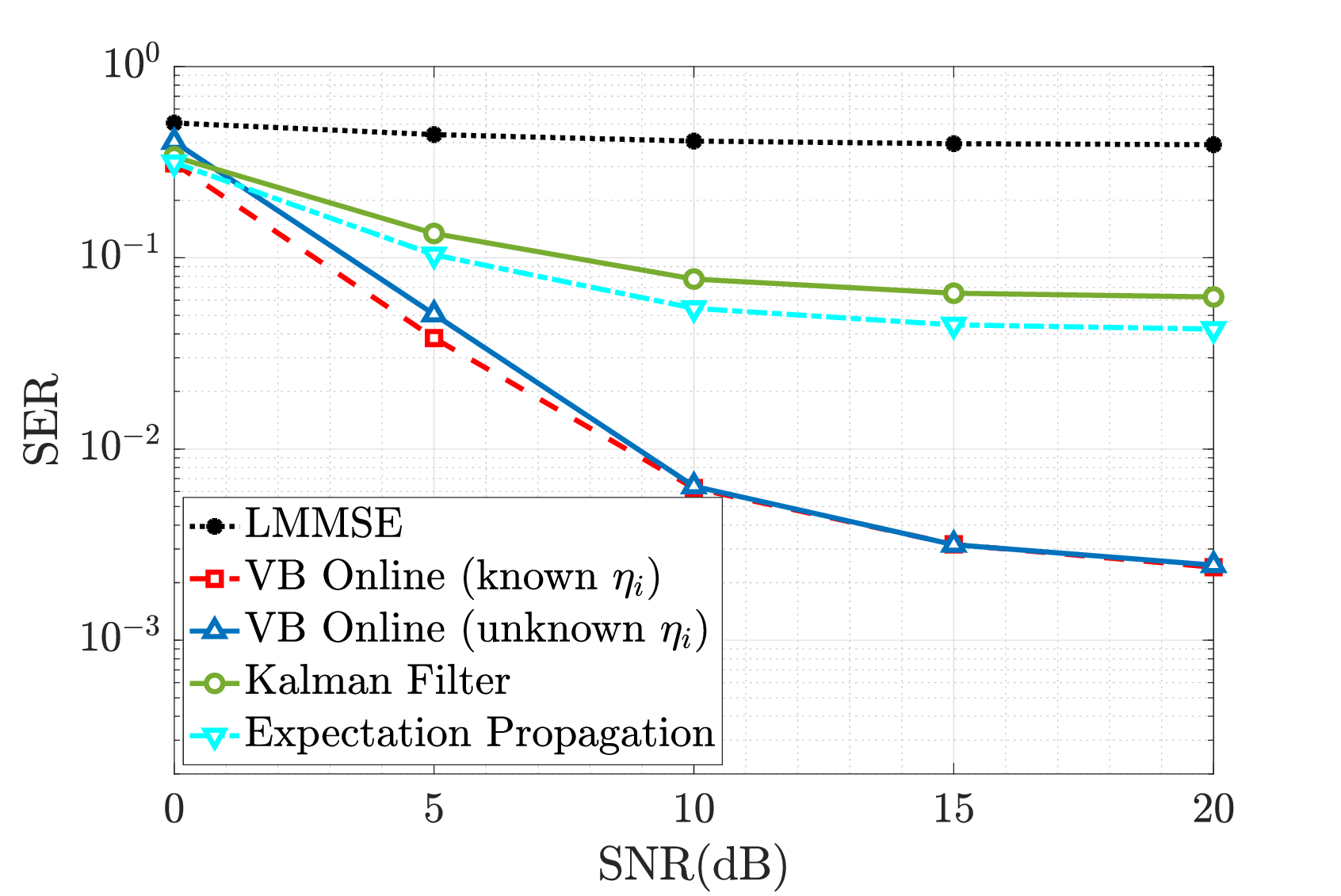}
  \vspace{-6mm}
  \caption{An SER comparison between LMMSE, KF, EP, and our online VB using QPSK under correlated channels when $\alpha = 0.5+j0.5, \eta_i= 0.985, T_p = 8, T_d=128$, and $\text{SNR}\in [0,20]$ dB.
  }\label{Fig:VB_benchmarks_QPSK_eta_fix}
   \vspace{-2mm}
\end{figure}

{\bf Case I -- $\bs{\eta_i}$ is fixed:} We assume correlated channels with $\alpha = 0.5+j0.5$ when $M = 32$, $K=4$, and $\eta_i = 0.985$, where $\eta_i$ is unknown at the BS. Notice that $\eta_i = 0.985$ signifies the Doppler frequency of the $i^{\mathrm{th}}$ high-mobility user traveling at a velocity of $158$ Km/h~\cite{naraghi2021semiICC}. Then, we let the VB framework estimate and track this parameter.  Fig.~\ref{Fig:VB_benchmarks_QPSK_eta_fix} illustrates the SER versus SNR using quadrature phase-shift keying (QPSK). As depicted in Fig.~\ref{Fig:VB_benchmarks_QPSK_eta_fix}, both VB-based approaches, with known and unknown $\eta_i$, exhibit nearly identical performance. This observation validates the efficacy of the proposed VB-based method in successfully tracking and estimating $\eta_i$. Further, despite the VB method with unknown $\eta_i$ estimating the parameter over the air, its performance surpasses that of the LMMSE, KF, and EP methods, while these benchmarks require knowing $\eta_i$ before communication. 
This is due to the fact that LMMSE does not take the time correlation into account, and the KF method is a one-shot approach, which only passes over the frame once, and thus cannot improve its performance during the process. Moreover, in Fig.~\ref{Fig:VB_benchmarks_QPSK_eta_fix}, our proposed online VB method demonstrates superior performance compared to the EP method because it treats $\gamma_t$ as an unknown variable, whereas the EP method in \cite{naraghi2021semiICC} assumes it is a fixed known component. This variable nature of $\gamma_t$ grants the VB framework greater flexibility in estimating $\bs{\Sigma}_{i,t}$ in \eqref{Sigma-h}, leading to its enhanced performance relative to the EP method.

\begin{figure}[t]
 \vspace{-3mm}
\centering
  \includegraphics[trim = 0mm 0mm 0mm 0mm, clip, scale=9, width=0.99\linewidth, draft=false]{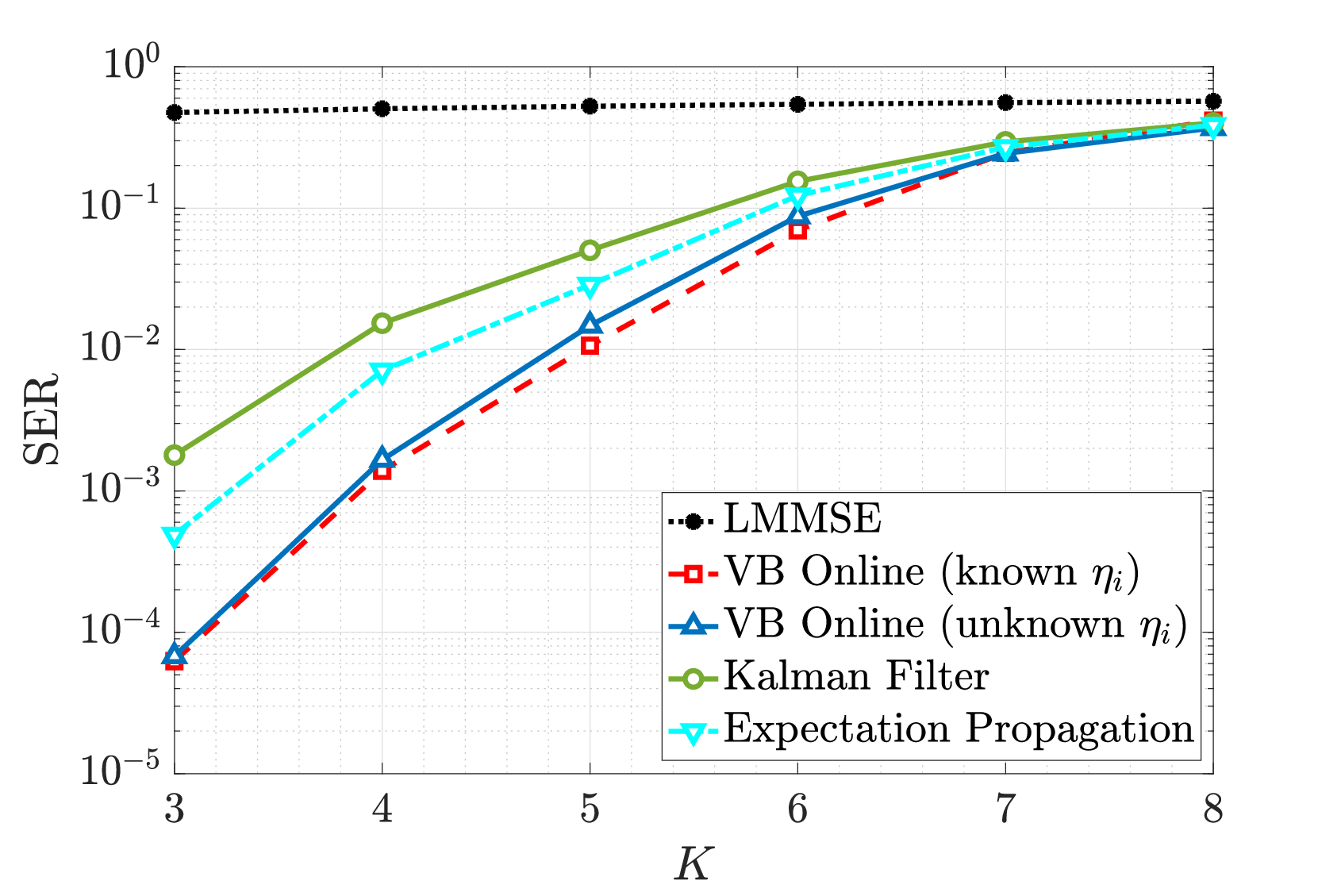}
  \vspace{-6mm}
  \caption{The SER performance comparison between LMMSE, KF, EP, and our online processing VB-based strategy utilizing QPSK modulation, when $\eta_i = 0.97$, $T_p = 8$, $T_d = 128$, $\text{SNR}=15$ dB, and $K \in [3,8]$.}\label{Fig:VB_benchmarks_QPSK_SER_diff_users}
  \vspace{-4mm}
\end{figure}

\begin{figure}[t]
\vspace{-3mm}
\centering
  \includegraphics[trim = 0mm 0mm 0mm 0mm, clip, scale=9, width=0.99\linewidth, draft=false]{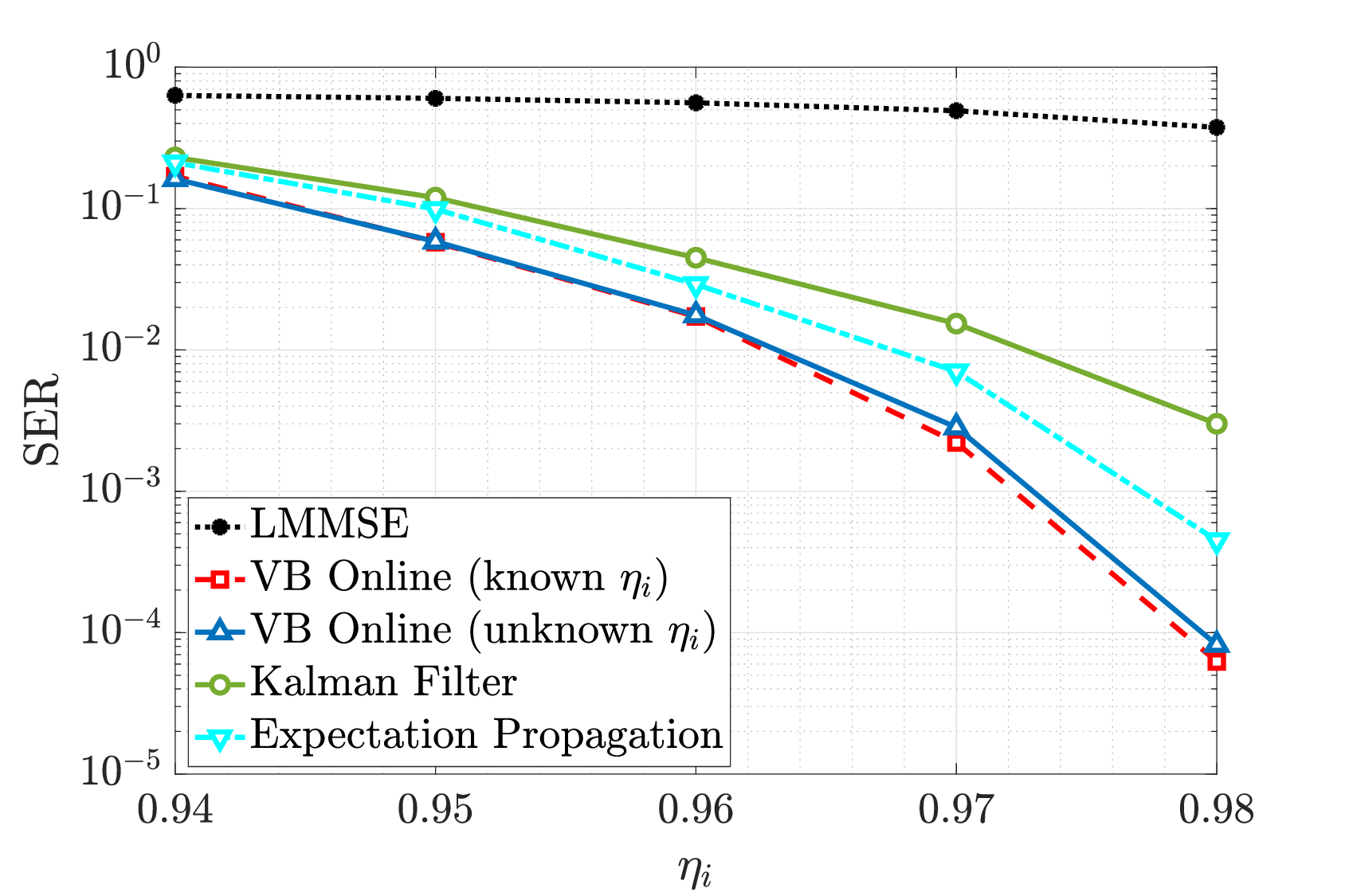}
  \vspace{-6mm}
  \caption{Comparison of SER performance for LMMSE, KF, EP, and the proposed online VB-based strategy using QPSK modulation, with $\eta_i \in[0.94, 0.98]$, $T_p = 8$, $T_d = 128$, and $\text{SNR}=15$ dB.}\label{Fig:VB_benchmarks_QPSK_diff_eta}
  \vspace{-4mm}
\end{figure}
Then, Fig.~\ref{Fig:VB_benchmarks_QPSK_SER_diff_users} illustrates the SER performance of our method in comparison to three benchmark approaches, with $M=32$, $\text{SNR}=15 ~ \mr{dB}$, $\eta_i = 0.97$, $\mb{R}_i = \frac{1}{M} \mb{I}_M$, and $K \in [3,8]$. It is evident that the performance of all methods declines as $K$ increases. This occurs as we can treat a network with $K_2$ users similarly to one with $K_1$ users, where $K_1 < K_2$, by setting $K_2 - K_1$ of the $K_2$ users to be inactive. However, when these $K_2 - K_1$ users become active, they contribute additional interference, further complicating channel estimation and data detection. This is exacerbated by the fact that the VB framework, which tends to converge to local optimums, cannot perfectly eliminate errors during the JED process. As a result, as the number of users increases, the probability of errors in estimation and detection rises, leading to a higher SER. Next, in Fig.~\ref{Fig:VB_benchmarks_QPSK_diff_eta}, we compare the performance of our online VB method with the LMMSE, KF, and EP methods for various values of $\eta_i \in [0.94, 0.98]$ when $M=32$, $K=4$, $\mb{R}_i = \frac{1}{M} \mb{I}_M$, and $\text{SNR}=15 ~ \mr{dB}$. The results indicate that the SER performance of all methods improves as $\eta_i$ increases. This happens because, as $\eta_i$ approaches $1$, the variation in the channels in the Gauss-Markov model in \eqref{GM-model} between consecutive time slots decreases. This leads to improved accuracy in channel estimation and, consequently, better SER performance.

\begin{figure}[t]
\centering
  \includegraphics[trim = 0mm 0mm 0mm 0mm, clip, scale=9, width=0.99\linewidth, draft=false]{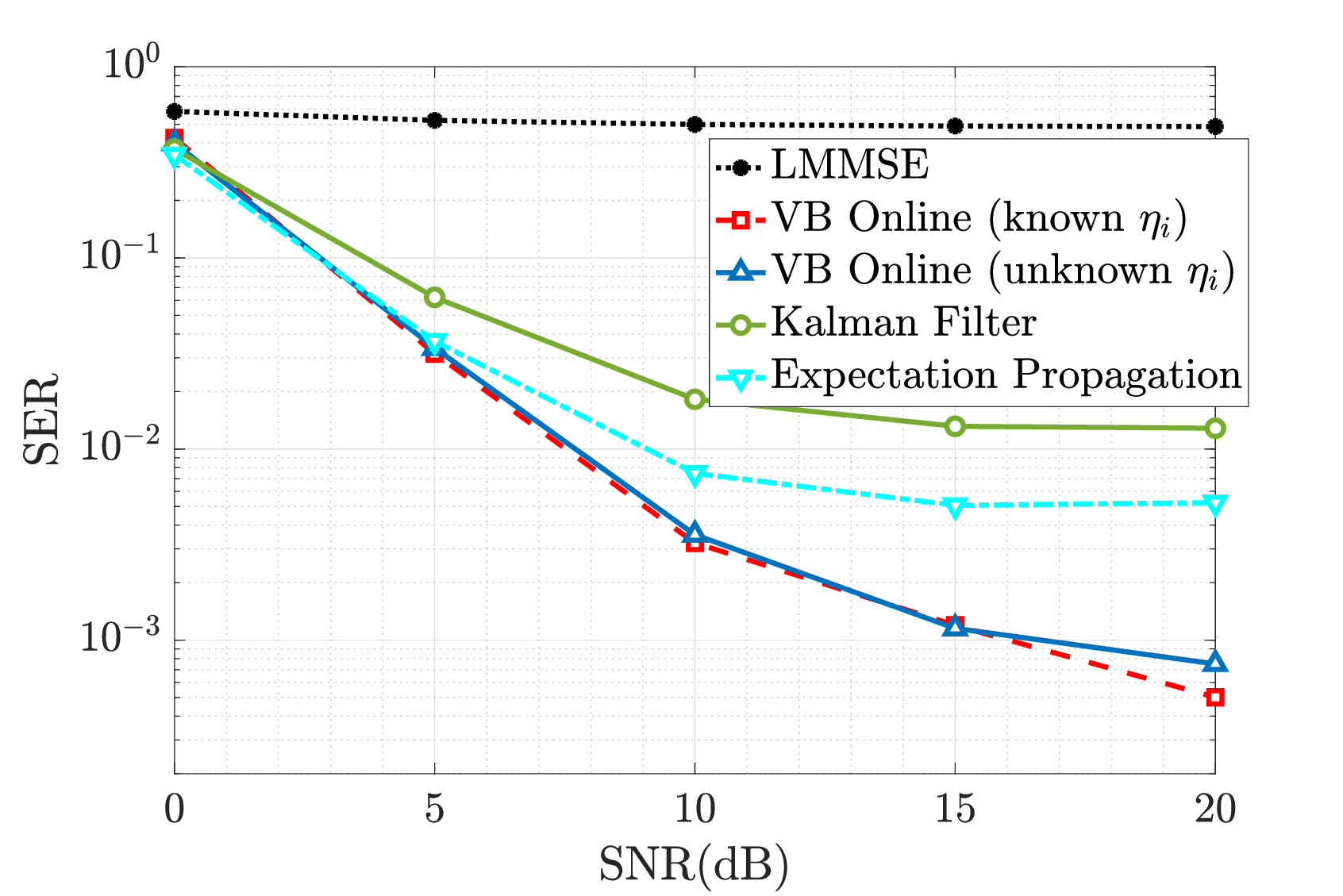}
  \vspace{-6mm}
  \caption{A comparison of SER among LMMSE, KF, EP, and our online processing VB-based approach employing QPSK modulation where $\eta_i \sim \mathcal{N}(0.97, 5 \times 10^{-5})$, $T_p = 8$, $T_d=128$, and SNR ranging from $0$ to $20$ dB.  }\label{Fig:VB_benchmarks_QPSK_eta_random}
  \vspace{-4mm}
\end{figure}

{\bf Case II -- $\bs{\eta_i}$ is a random variable:} To cover a more realistic scenario, we consider a situation where $\eta_i$ is a random variable with minor fluctuations. This assumption is motivated by the fact that the Doppler frequency of the $i^{\mathrm{th}}$ user could change slowly over the communication time due to changes in its velocity and direction. Thus, in this context, we model $\eta_i \sim\mathcal{N}(0.97, 5 \times 10^{-5})$, while $M = 32$, $K=4$, $\mb{R}_i = \frac{1}{M} \mb{I}_M$, and $\eta_i = 0.97$. As illustrated in Fig.~\ref{Fig:VB_benchmarks_QPSK_eta_random}, the observed trend closely mirrors that of the scenario where $\eta_i$ is fixed, indicating the effectiveness of the proposed online processing approach in accurately estimating and tracking the variable correlation coefficient $\eta_i$, consequently showcasing superior performance over the baseline methods in terms of SER. 

\begin{figure}[ht]
\vspace{-3mm}
\centering
  \includegraphics[trim = 0mm 0mm 0mm 0mm, clip, scale=9, width=0.99\linewidth, draft=false]{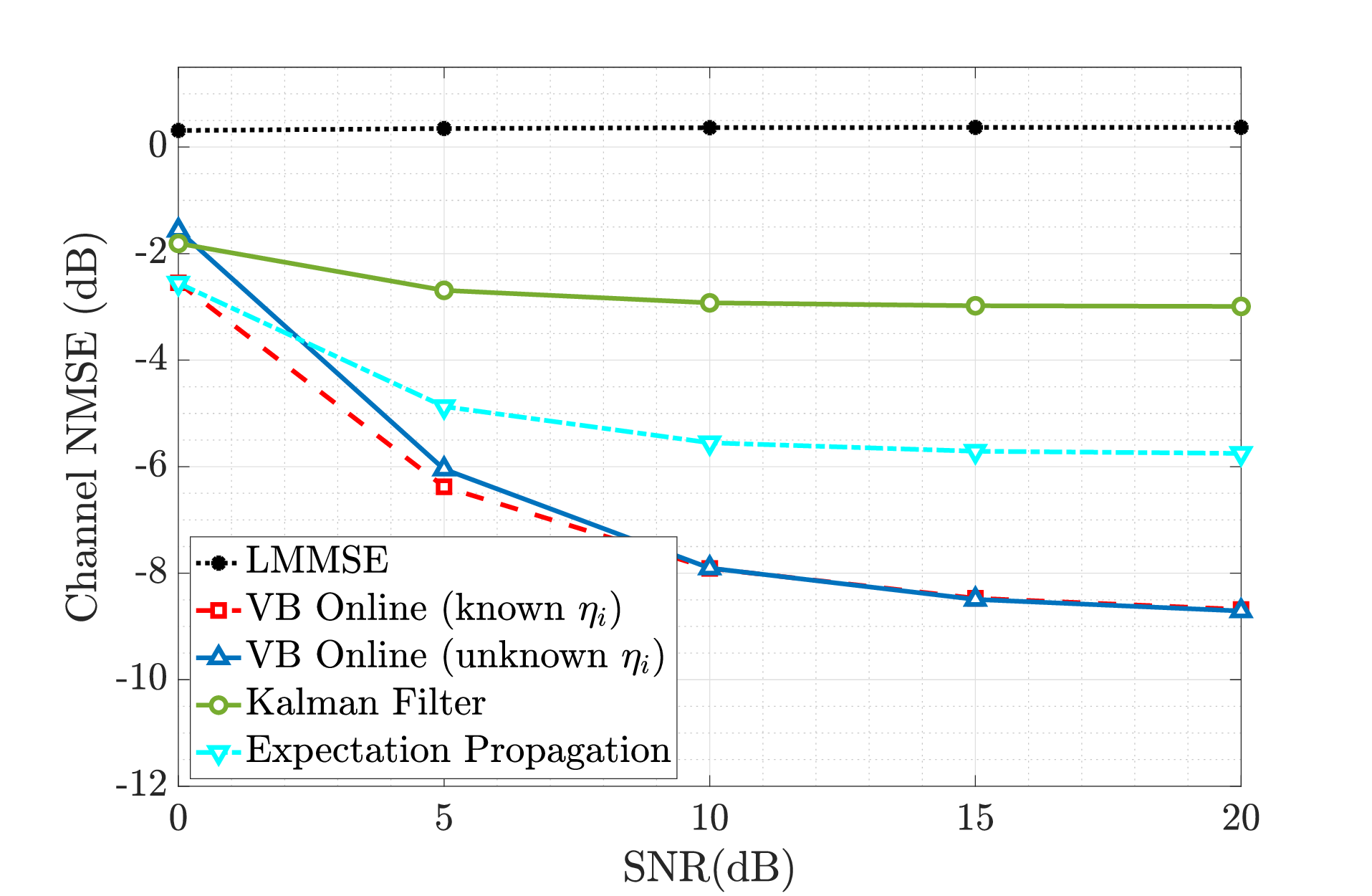}
  \vspace{-6mm}
  \caption{The channel NMSE performance comparison between LMMSE, KF, EP, and our online VB utilizing QPSK modulation under correlated channels, where the parameters are set to $\alpha = 0.5+j0.5, \eta_i = 0.985$, $T_p = 8$, $T_d = 128$, and $\text{SNR}\in [0,20]$ dB.}\label{Fig:VB_benchmarks_QPSK_eta_fix_MSE}
  \vspace{-4mm}
\end{figure}

Secondly, since the channel estimation process plays a crucial role in system performance, we study the channel NMSE performance of our online processing method alongside other benchmarks here. The simulation setup mirrors that depicted in Fig.~\ref{Fig:VB_benchmarks_QPSK_eta_fix}. According to Fig.~\ref{Fig:VB_benchmarks_QPSK_eta_fix_MSE}, our proposed online method, based on VB, yields the lowest NMSE results compared to the baselines. This observation elucidates that the superior channel NMSE performance of our method comes from its more precise data detection process compared to the other benchmarks. Furthermore, as shown in Fig.~\ref{Fig:VB_benchmarks_QPSK_eta_fix_MSE}, all methods exhibit an error floor. This phenomenon arises from two primary sources of error in the channel estimation process: (1) weak signals relative to noise, which occurs at low SNRs, and (2) the time-varying nature of the channels. The figure demonstrates that increasing SNR (i.e., receiving a stronger signal) improves estimation performance up to $\text{SNR} = 10$ dB with the online VB method. Beyond this point, however, further improvement is minimal because the channels continue to vary over time, making it difficult for the approximated statistical model of the channel at one time slot to remain accurate in the next time slot, even with stronger signals.

Finally, it is crucial to assess the computational complexity of our approach in comparison to the benchmarks. The computational complexity of LMMSE is $\mathcal{O}\left(MK^2+|\mathcal{S}|K\right)$, while for both KF and EP methods, it is $\mathcal{O}\left(I_{\mathrm{tr}}\left(M^3K^3 + M^2 K^2 + |\mathcal{S}| K\right)\right)$~\cite{naraghi2021semi}. However, our online processing VB strategy has a computational complexity of $\mc{O}(I_{\mr{tr}}(M^3 K+|S|K))$. This complexity corresponds to finding a local optimum solution using the CAVI algorithm within the VB method~\cite{nguyen2022variational}. As a result, our online processing strategy within the VB framework has lower computational complexity than the KF and EP methods and higher complexity than the LMMSE method.

\begin{figure}[t]
\centering
  \includegraphics[trim = 0mm 0mm 0mm 0mm, clip, scale=9, width=0.99\linewidth, draft=false]{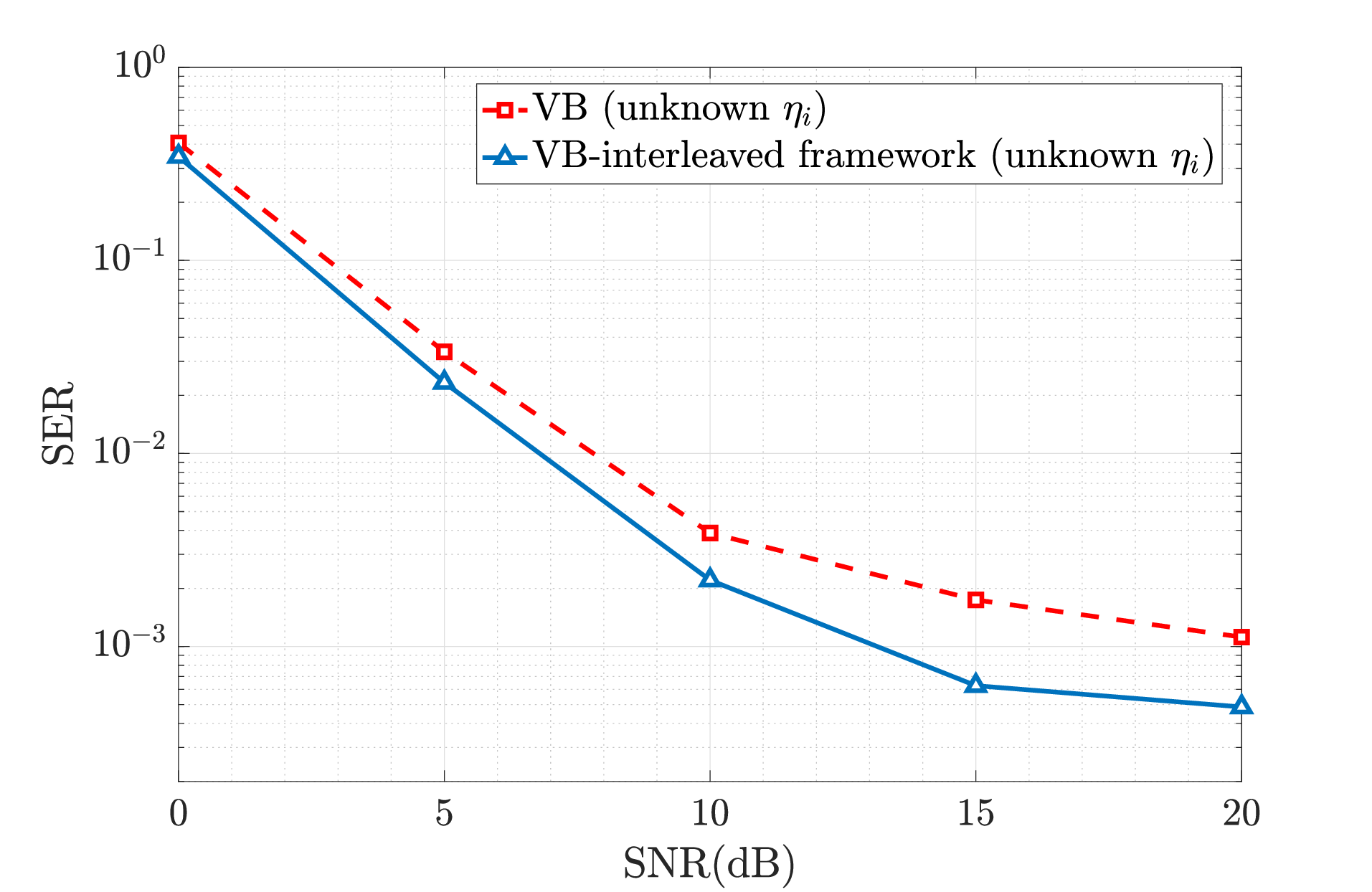}
  \vspace{-6mm}
  \caption{The SER performance of the proposed online processing using the VB framework with and without interleaved structure under the QPSK signal constellation when $\eta_i =0.97, T_p = 8, T_d=128$, and $\text{SNR}\in [0,20]$ dB.
  }\label{Fig:VB_intrlv}
  \vspace{-5mm}
\end{figure}

\begin{figure*}[t]
\vspace{-5mm}
\centering
  \includegraphics[trim = 0mm 0mm 0mm 0mm, clip, scale=7, width=0.9\linewidth, draft=false]{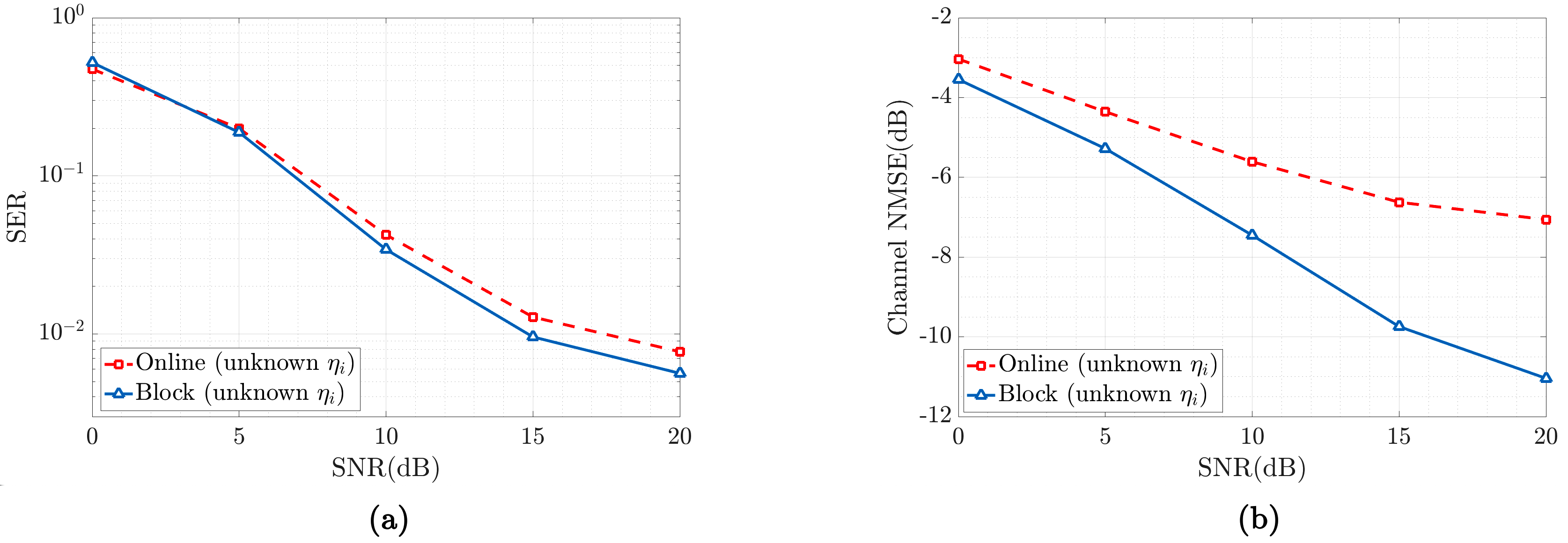}
  \vspace{-3mm}
  \caption{The SER and channel NMSE performance of the proposed online and block processing strategies using the VB framework with 16QAM when $M= 64$, $\eta_i =0.985, T_p = 8, T_d=128$, and $\text{SNR}\in [0,20]$ dB.
  }\label{Fig:VB_online_vs_block}
  \vspace{-5mm}
\end{figure*}

\subsection{Enhancing the Performance of the Online VB Strategy}

As mentioned earlier in Section~\ref{Sec:VB_for_JED}, the online processing strategy involves two phases of prediction and estimation, where the channel information at time $t-1$ serves as prior knowledge for time $t$. However, if the predicted channels at time $t-1$ fail to provide accurate prior knowledge for estimating the channel at time $t$, it can lead to error propagation in the estimation process. More precisely, by increasing $T_p$ and $T_d$, although we expect to receive more signals and thus improve the SER performance, the predicted channels from the prediction phase become more and more deviated from the actual channel due to the time correlation between the channels. Therefore, increasing $T_p$ and $T_d$ alone cannot lead to performance improvement. To address this issue, we propose an interleaved structure that divides the entire communication time into $L$ sections, each with its own pilot and data signals. This way, our VB-based method leverages the interleaved structure to stop the error propagation and re-tune the estimated parameters based on the new pilot signals. Here, we have:
\begin{align}
\sum_{\ell = 1}^L T_p^{[\ell]} = T_p, \quad \sum_{\ell = 1}^L T_d^{[\ell]} = T_d,
\end{align}
where $T_p^{[\ell]}$ and $T_d^{[\ell]}$ represent the pilot and data communication times for the $\ell^{\mathrm{th}}$ section, respectively.

In Fig.~\ref{Fig:VB_intrlv}, we incorporate an interleaved structure with $L=2$ sections into the online processing strategy and compare its performance with a scenario where the structure is not utilized. We employ the same assumptions as in Fig.~\ref{Fig:VB_benchmarks_QPSK_eta_fix}. The results show that the interleaved structure enhances the SER compared to the non-interleaved scenario. This improvement arises because the interleaved structure prevents error propagation and enables the BS to adjust parameters using new pilot signals at the beginning of each interleaving section.

\subsection{Online Strategy versus Block Processing Strategy}
In Fig.~\ref{Fig:VB_online_vs_block}, we compare the performance of the block processing strategy with the online processing strategy in terms of SER and channel NMSE. We assume $M = 64$ and $\eta_i = 0.985$ using the 16-quadrature amplitude modulation (16QAM) signal constellation. Fig.~\ref{Fig:VB_online_vs_block} depicts that the block processing strategy slightly enhances SER performance. Moreover, it significantly reduces channel NMSE, especially at high SNRs. These improvements stem from stacking the entire received signals together and processing them together. On the other hand, the block processing strategy increases delay and computational complexity compared to the online processing strategy. In particular, the complexity of the block processing VB method is expressed as 
$\mc{O}(I_{\mr{tr}}\big[T(3M^2 + M^3)K + M^3 K + |\mc{S}|K\big])$, due to the matrix inversion in \eqref{Sigma-h_block} and the matrix multiplications required to compute 
$\tau^{\eta}_{i,t}$, $\lr{\eta_{i,t}}$, and 
$\bar{b}_{i,t}$ in \eqref{tau_i_var_block}, \eqref{tau_i_mean_block}, and \eqref{estimation-b-i_block}, respectively. As we can see, this complexity exceeds that of the online VB method, which explains the trade-off between higher complexity and enhanced performance in the block processing strategy.

To highlight the applications of the online and block processing strategies, we note that the online strategy is useful in latency-sensitive scenarios, such as voice and video calls, as well as in environments with limited computational resources, such as IoT networks. In contrast, block processing is preferable when lower error rates are required (e.g., for control signaling communication in cellular networks) and sufficient computational power is available.

%% file: Sections/Conclusion.tex
In this study, we focused on the uplink scenario within a massive MIMO network. We introduced approaches based on VB inference to tackle the JED problem, particularly for high-mobility users experiencing time-varying channels. We developed two processing strategies using VB: (i) a low-complexity low-latency online processing strategy suitable for scenarios with unknown time correlation coefficients at the BS; (ii) a block processing strategy that further enhances performance by reducing channel NMSE. Moreover, we incorporated an interleaved structure into the online processing strategy to prevent error propagation during the JED process, thereby improving its effectiveness. We numerically compared the performance of the VB-based online strategy against three well-known benchmarks (LMMSE, KF, and EP) in terms of SER and channel NMSE. Our simulation results consistently demonstrated the superiority of our VB framework across the performance metrics. One potential direction for future exploration involves extending these findings to networks integrating reconfigurable intelligent surfaces (RISs) with massive MIMO systems, considering RIS's emerging role as a promising, low-complexity, and power-efficient solution in 5G and beyond networks~\cite{mei2021multi,nassirpour2023beamforming}.